\documentclass[12pt,preprint]{aastex}

\shorttitle{Resolving the nuclear region of z=5.78 quasar hosts}
\shortauthors{Wang et al.}

\begin{document}


\title{Resolving the interstellar medium in the nuclear region of 
two z=5.78 quasar host galaxies with ALMA}


\author{Ran Wang\altaffilmark{1}, 
Yali Shao\altaffilmark{1,2}, 
Chris L. Carilli\altaffilmark{3,4}, 
Gareth C. Jones\altaffilmark{3,5}, 
Fabian Walter\altaffilmark{6,7}, 
Xiaohui Fan\altaffilmark{8}, 
Dominik A. Riechers\altaffilmark{9}, 
Roberto Decarli\altaffilmark{6,10},
Frank Bertoldi\altaffilmark{11}, 
Jeff Wagg\altaffilmark{12}, 
Michael A. Strauss\altaffilmark{13}, 
Alain Omont\altaffilmark{14}, 
Pierre Cox\altaffilmark{14}, 
Linhua Jiang\altaffilmark{1}, 
Desika Narayanan\altaffilmark{15,16,17}, 
Karl M. Menten\altaffilmark{18},
Bram P. Venemans\altaffilmark{6}
}
\altaffiltext{1}{Kavli Institute of Astronomy and Astrophysics at Peking University, 
No.5 Yiheyuan Road, Haidian District, Beijing, 100871, China}
\altaffiltext{2}{Department of Astronomy, School of Physics, Peking University, Beijing 100871, China}
\altaffiltext{3}{Cavendish Laboratory, University of Cambridge, 19 J. J. Thomson Avenue, Cambridge CB3 0HE, UK}
\altaffiltext{4}{National Radio Astronomy Observatory, Socorro, NM 87801-0387, USA}
\altaffiltext{5}{Kavli Institute for Cosmology, University of Cambridge, Madingley Road, Cambridge CB3 0HA, UK}
\altaffiltext{6}{Max-Planck-Institut for Astronomie, K\"{o}nigstuhl 17, D-69117 Heidelberg, Germany}
\altaffiltext{7}{Astronomy Department, California Institute of Technology, MC105-24, Pasadena, CA 91125, USA}
\altaffiltext{8}{Steward Observatory, University of Arizona, 933 North Cherry Avenue, Tucson, AZ 85721, USA}
\altaffiltext{9}{Department of Astronomy, Cornell University, 220 Space Sciences Building, Ithaca, NY 14853, USA}
\altaffiltext{10}{INAF—Osservatorio di Astrofisica e Scienza dello Spazio, via Gobetti 93/3, I-40129, Bologna, Italy}
\altaffiltext{11}{Argelander-Institut f\"{u}r Astronomie, University at Bonn, Auf dem H\"{u}gel 71, D-53121 Bonn, Germany}
\altaffiltext{12}{SKA Organization, Lower Withington Macclesfield, Cheshire SK11 9DL, UK}
\altaffiltext{13}{Department of Astrophysical Sciences, Princeton University, Princeton, NJ 08544, USA}
\altaffiltext{14}{Institut d'Astrophysique de Paris, CNRS and Universite Pierre et Marie Curie, 98b boulevard Arago, 75014 Paris, France}
\altaffiltext{15}{Department of Astronomy, University of Florida, 211 Bryant Space Sciences Center, Gainesville, FL 32611 USA}
\altaffiltext{16}{University of Florida Informatics Institute, 432 Newell Drive, CISE Bldg E251, Gainesville, FL 32611}
\altaffiltext{17}{Cosmic Dawn Center (DAWN), Niels Bohr Institute, University of Copenhagen, Juliane Maries vej 30, DK-2100 Copenhagen, Denmark}
\altaffiltext{18}{Max-Planck-Institut f{\"u}r Radioastronomie, Auf dem H\"{u}gel 69, 53121 Bonn, Germany}

\begin{abstract}

We present ALMA observations of the [CII] 158 $\mu$m fine structure line and dust 
continuum emission from two quasars,  SDSS J104433.04-012502.2 and SDSS J012958.51-003539.7, 
at z=5.78. The ALMA observations at 0.2$''$ resolution map the dust and gas on kpc scales. The 
spatially resolved emission shows a similar trend of decreasing [CII]-FIR ratios with increasing 
FIR surface brightnesses as was found in the infrared luminous 
galaxies with intense star formation. We confirm the velocity gradients of [CII] emission 
found previously in SDSS J0129-0035. No clear evidence of order motion is detected in SDSS J1044-0125. 
The velocity maps and PV-diagrams also suggest turbulent gas clumps in both objects. 
We tentatively detect a [CII] peak offset 4.9 kpc to the East of SDSS J1044-0125. This may 
be associated with an infalling companion, or node of gas outflow. All these results suggest 
significant dynamical evolution of the ISM in the nuclear region of these young quasar-starburst 
systems. We fit the velocity map of the [CII] emission from SDSS J0129-0035 with a rotating 
disk model. The result suggests a face-on system with an inclination angle of  $\rm 16\pm20^{\circ}$ 
and constrains the lower limit of the host galaxy dynamical 
mass to be $\rm 2.6\times10^{10}\,M_{\odot}$ within the [CII] 
emitting region. It is likely that SDSS J0129-0035, as well as 
other young quasars with super  massive black hole masses on 
order of $\rm 10^{7}\,M_{\odot}$  to $\rm 10^{8}\,M_{\odot}$, falls close 
to the black hole and host galaxy mass relation defined by local galaxies.

\end{abstract}





\section{Introduction}

Quasars at high redshifts provide the best opportunity to study the formation and early evolution 
of the first supermassive black holes (SMBHs) and their host galaxies. Due to the great success of 
wide-field optical and near-infrared surveys in the last two decades, more than two hundred  
quasars have been discovered at z$\geq$5.7, an epoch close to the end of cosmic reionization 
(e.g. \citealp{fan06,mortlock11,venemans13,banados16,banados18,jiang09,jiang16,matsuoka16,
matsuoka18a,matsuoka18b,chehade18}). 
This quasar sample covers a wide range of SMBH masses and quasar luminosities, including the most massive sources 
with SMBH mass above $\rm 10^{10}\,M_{\odot}$ (e.g., \citealp{fan00,wu15}) as well as 
objects with SMBH masses of a few $\rm 10^{7}$ 
to $\rm 10^{8}\,M_{\odot}$ \citep{willott10,jiang09,matsuoka16,matsuoka18a,matsuoka18b} 
which are more similar to the common quasar population observed at lower redshifts.  

The dust and gas in the host galaxies of these earliest quasars have been studied at 
submillimeter and millimeter wavelengths using thermal continuum emission, molecular (mostly CO), 
and fine structure lines, such as the [C II] 158$\mu$m emission line \citep{walter04,walter09,bertoldi03a,bertoldi03b,petric03,priddey03,robson04,riechers09,omont13,
wang08,wang10,wang13,willott15,willott17,banados15,venemans16,venemans17a,venemans17b,
venemans17c,decarli17,decarli18,izumi18}. Dust continuum surveys at $\sim$0.5-1 mJy rms  
sensitivity have detected the most FIR-luminous objects, with far-infrared (FIR) luminosities of a few 
$\rm \times10^{12}$ to $\rm 10^{13}\,L_{\odot}$ and dust masses of $\rm >10^{8}\,M_{\odot}$ 
\citep{bertoldi03a,priddey03,robson04,wang07,wang08}. 
These FIR luminous quasars are bright in CO line emission and [C II] 158$\mu$m 
fine structure line emission, indicating molecular gas with masses of order of $\rm 10^{10}\,M_{\odot}$ 
and bursts of star formation in the nuclear region, co-eval with the SMBH accretion and quasar 
activity \citep{bertoldi03b,maiolino05,carilli07,wang13,decarli17,decarli18,venemans16}.  

In recent years, the Atacama Large Millimeter/submillimeter Array (ALMA) has carried out comprehensive 
surveys of the [C II] 158$\mu$m fine structure line in high-z quasars. For example, \citet{decarli18} 
detected [C II] in 85\% of 27 optically-selected quasars at z$>$5.94. 
This line is an important coolant that traces the ionized and neutral interstellar 
medium (ISM) and star forming activities \citep{herreracamus15}. 
The detection of [C II] line and dust continuum emission has revealed a wide range of star formation 
rates, from a few 10 $\rm M_{\odot}\,yr^{-1}$ to $\rm \geq1000\,M_{\odot}\,yr^{-1}$ 
\citep{decarli18,izumi18}.  
The [C II]-FIR luminosity ratios of these quasar hosts range from $\rm 10^{-4}$ to a few $\rm \times10^{-3}$ 
\citep{maiolino05,wang13,willott17,decarli18}, following the trend of 
decreasing [C II]-FIR ratio with increasing FIR luminosities found with the IR luminous star-forming 
systems \citep{malhotra01,stacey10,luhman03,diazsantos13,diazsantos17,gracia11,
haileydunsheath10,munoz16,smith17,gullberg18,decarli18,rybak19}. 
In particular, the quasar hosts with high FIR luminosities of a few $\rm 10^{12}$ 
to $\rm 10^{13}\,L_{\odot}$ show low [C II]-FIR ratios similar to that found in 
the Ultraluminous infrared galaxies (ULIRGs) and submillimeter 
galaxies (SMGs) \citep{luhman03,diazsantos13,wang13,diazsantos17,rybak19}.  
The [C II]-FIR ratio is suggested to be related to the local conditions of the 
ISM; e.g. a decrease of [C II]-FIR ratio could be due to a high gas 
temperature (i.e., $\rm T>\Delta E/k\sim 91\,K$, $\rm \Delta$E is the 
energy separation of the two levels of the [C II] 158 $\mu$m transition) where 
the upper level of [C II] transition is saturated \citep{munoz16}, or a high gas surface 
density (e.g., the typical density in compact ULIRGs and SMGs) where more gas is in molecular phase \citep{narayanan17}. 
It is interesting to see how the resolved distributions of the line and 
continuum surface brightnesses and emission ratios in these FIR luminous quasar hosts 
compare to that in the ULIRGs and SMGs \citep{smith17,gullberg18,rybak19}.

The [C II] 158$\mu$m and molecular CO lines are bright tracers of the gas content that could be detected from distant galaxies \citep{solomon05,cc13}. 
Measurements of the line widths, profiles, and velocity maps of these emission lines reveal important information on the gas 
kinematics \citep{walter04,ivison11,maiolino12,wang13,venemans16,venemans19}.
Observations of the [C II] and CO emission lines from the  
young quasar hosts at z$\geq$5.7 at sub-arcsecond to arcsecond resolution show a range of kinematic properties, 
including velocity gradients of ordered motion \citep{willott13,willott17,venemans16,shao17,feruglio18}, large velocity 
dispersion/turbulence (e.g., SDSS J231038.88+185519.7, \citealp{feruglio18}), 
gas outflows (e.g., SDSS J114816.64+515250.3, \citealp{maiolino12,cicone15}), 
and very compact sources with no evidence of rotation (e.g., ULAS J112001.48+064124.3, \citealp{venemans17c}).
All these submillimeter/millimeter (submm/mm) observations argue for an early phase of SMBH-galaxy co-evolution in 
the z$\geq$5.7 quasar sample. Deep imaging of the gas component at kpc 
scale is required to fully resolve the gas distribution and kinematic properties in 
these systems.  

In this paper, we present the observations, results and conclusion 
for two out of three quasars from our ALMA Cycle 1 program.
Our team has carried out ALMA observations at $\rm 0''.2-0''.3$ angular resolution to image the [C II] line emission 
from the host galaxies of three quasars, ULAS J131911.29+095051.4 (hereafter ULAS J1319+0950), 
SDSS J104433.04-012502.2 (hereafter SDSS J1044-0125), and SDSS J012958.51-003539.7 (hereafter SDSS J0129-0035) 
at z$\geq$5.7. These three objects are among the brightest dust continuum, and [C II] 
detections at z$\sim$6 \citep{wang13}, with FIR luminosities 
of $\rm 5\sim10\times10^{12}\,L_{\odot}$ \footnote{We adopt the rest-frame wavelength 
range of 42.5$\mu$m to 122.5$\mu$m for FIR luminosity calculation} and [C II] 
luminosities of $\rm 1.5\sim4.4\times10^{9}\,L_{\odot}$ \citep{wang13}, suggesting massive star 
formation in the quasar hosts. The three objects 
are different in quasar UV luminosities and SMBH masses. SDSS J1044-0125 and ULAS J1319+0950 are among 
the most optically luminous quasars at z$\sim$6 with rest-frame 1450$\AA$ apparent magnitudes of $\rm m_{1450}$=19.21   
and 19.65, respectively \citep{fan00,mortlock09}. 
The SMBH mass is $\rm (2.7\pm0.6)\times10^{9}\,M_{\odot}$ for ULAS J1319+0950 \citep{shao17} and 
$\rm (5.6\pm0.6)\times10^{9}\,M_{\odot}$ for SDSS J1044-0125 \citep{shen19}. 
SDSS J0129-0035 is much fainter in quasar emission with $\rm m_{1450}$=22.28 \citep{jiang09}. 
If Eddington accretion is assumed \citep{willott10}, the SMBH mass is $\rm 1.7\times10^{8}\,M_{\odot}$.
These objects provide a ideal sample for a pilot study of the ISM distribution and dynamics in 
the earliest SMBH-starburst systems over a range of quasar luminosities.

We report our ALMA [C II] line imaging of ULAS J1319+0950 at $\rm 0.3''$ resolution in \citet{shao17}.  
The [C II] velocity map of this object shows clear velocity gradients and the spectrum appears 
to be broad ($\rm FWHM\sim540\,km\,s^{-1}$) and flat at the top,
suggesting that the [C II]-emitting gas is mainly distributed in an inclined rotating disk
extending to a radius of $\sim$3.2 kpc. We performed dynamical modeling of the gas velocity field and measured
the disk inclination angle and rotation velocity of the atomic/ionized gas in the nuclear region.
The derived host galaxy dynamical mass is $\rm 1.3\times10^{11}\,M_{\odot}$, suggesting a black hole-to-host galaxy mass
ratio of $\sim0.02$, i.e., about four times higher than the value expected from the black hole-bulge mass relation of local
galaxies \citep{kormendy13}.

We here present ALMA [C II] imaging of the remaining two quasars, SDSS J0129-0035 and SDSS J1044-0125.  
We describe the observations in Section 2, present the results in Section 3, discuss the surface brightnesses of the [C II] 
and dust emission and gas kinematic  
properties in Section 4, and summarize the main conclusions in Section 5. 
We adopt a $\Lambda$CDM cosmology with $\rm H_{0}=71km\,s^{-1}\,Mpc^{-1}$,
$\rm \Omega_{M}=0.27$ and $\rm \Omega_{\Lambda}=0.73$ throughout this paper \citep{spergel07}.

\section{Observations}

We carried out observations of the [C II] 158 $\mu$m emission line from 
SDSS J0129$-$0035 and SDSS J1044$-$0125 using the Band 7 receiver on ALMA in Cycle 1 (Program ID: 2012.1.00240.S). 
The data was taken in 2015 July and August in the C34-6 configuration 
with 32 12-m diameter antennas and a maximum baseline of 1.6 km. 
This results in a full width at half maximum (FWHM) synthesized beam size of $\sim$0.2$''$ using robust weighting, 
which corresponds to 1.2 kpc at z$\sim$5.8.
We tuned the four 1.785 GHz spectral windows with one window centered on the redshifted [C II] line frequencies 
of the two targets and the other three windows observing the continuum.
The correlator channel width is 15.625 MHz, corresponding to a velocity resolution 
of $\sim$17 $\rm km\,s^{-1}$. We checked the phase every 5$\sim$7 minutes on nearby 
calibrators, and the flux scale was calibrated on Ceres and J1058+015 with calibration 
uncertainties less than 10\%. The total on-source time is about 80 min for SDSS J0129-0035 and 75 min for SDSS J1044-0125. 

The data were calibrated using the pipeline of the Common Astronomy Software Application (CASA) 
Version 4.3.1. We then combined the data with that obtained from previous ALMA Cycle 0 
observations (Program ID: 2011.0.00206.S, \citealp{wang13}). 
The ALMA Cycle 0 data alone has a lower angular resolution of 0.6$''$-0.7$''$, and 1$\sigma$ rms 
sensitivities of 0.35 mJy $\rm beam^{-1}$ for SDSS J0129-0035 
and 0.65 mJy $\rm beam^{-1}$ for SDSS J1044-0125 in each 67 $\rm km\,s^{-1}$ channel \citep{wang13}.  
By including the Cycle 0 data, we could better sample the uv plane at different scale and improve 
the sensitivity at short uv-distance. 
As the Cycle 0 data were calibrated using a much earlier 
version of CASA, we corrected the weights before combining with the Cycle 1 
data\footnote{https://casaguides.nrao.edu/index.php/DataWeightsAndCombination}.
We checked the amplitudes of visibility data from the two observations at overlapping uv-distance, 
and the flux scales are consistent with each other within the calibration uncertainties.
We averaged, in the visibility domain, the emission in line-free channels to form a 'pseudo' 
continuum dataset which we then subtracted from the uv-data to obtain a line-only dataset.
We made the continuum image with the data from three line-free spectral windows and made the line image 
cubes with the continuum-subtracted visibility dataset.
The images were cleaned using the CLEAN task in CASA using robust=0.5 for weighting. 
The typical rms noise is 0.20 mJy $\rm beam^{-1}$ per 17 $\rm km\,s^{-1}$ channel  
for the final image cube of SDSS J0129$-$0035, and 0.36 mJy $\rm beam^{-1}$ per 17 $\rm km\,s^{-1}$ channel 
for SDSS J1044$-$0125. The rms noise of the continuum is 21$\rm \mu Jy\,beam^{-1}$ for SDSS J0129$-$0035, and  
40$\rm \mu Jy\,beam^{-1}$ for SDSS J1044$-$0125. 

\section{Results}

For each of the two objects, we integrated the [C II] emission over the line-emitting
channels to obtain velocity-integrated maps and averaged the line-free channels to 
obtain the continuum maps (Figure 1). 
We then integrated the emission within the 2$\sigma$ contour region to get total [C II] line
fluxes and continuum flux densities (Table 1). For SDSS J0129-0035, the [C II] velocity-integrated map 
has a peak of $\rm 0.80\pm0.03\,Jy\,beam^{-1}\,km\,s^{-1}$ 
(signal-to-noise ratio, $\rm SNR=23.5$), which is about 38\% of the total line flux. 
The continuum intensity map has a peak of $\rm 1.58\pm0.02\,mJy\,beam^{-1}$ ($\rm SNR=80$), 
56\% of the integrated continuum flux density. For SDSS J1044-025, 
the [C II] peak is $\rm 0.64\pm0.08\,Jy\,beam^{-1}\,km\,s^{-1}$ (i.e., $\rm SNR=8.4$). 
This is about 45\% of the total line flux. The peak of the continuum emission
is $\rm 1.88\pm0.04\,mJy\,beam^{-1}$ ($\rm SNR=47.5$), which is about 63\% of the 
total continuum flux density. Thus, we consider the [C II] and continuum emission 
from both objects are resolved. 

The visibilities versus uv-distance for the [C II] line and continuum are plotted in Figure 2. 
The data are radially averaged with a bin size of 50 k$\lambda$ for SDSS J0129-0035 
and 75 k$\lambda$ for SDSS J1044-0125. The decrease of the visibility amplitudes 
toward longer uv-distance indicates that the line and continuum sources are 
resolved in the uv-plane. We performed elliptical Gaussian fitting to measure
the sizes of the [C II] and continuum emission in both the image plane and the uv
plane, using {\em imfit} and {\em uvmodelfit} tasks in CASA, respectively. 
The results, together with the integrated line fluxes and continuum flux densities, 
are summarized in Table 1. The uv-models are also plotted in Figure 2. 
The Gaussian FWHM source sizes (deconvolved from the synthesized
beam) and peak positions in the image plane are consistent with that from the uv-models within the 
fitting uncertainties. We adopt the fitting positions and deconvolved source sizes 
from {\em imfit} for the discussions in the rest of this paper.

The fitting uncertainties of the [C II] peak positions 
are $\sim$7 mas in both RA and DEC for SDSS J0129-0035, 
and 25 mas for SDSS J1044-0125. The positional uncertainties for the continuum images are $<$3 mas for both objects. 
Thus, the peak positions of the [C II] line and continuum emission are consistent with each other 
within the fitting errors. We checked the coordinates of the optical quasars using the astrometric data from 
the Gaia mission \citep{lindegren18}. For SDSS J0129-0035, the quasar coordinates 
are RA=01$^{h}$29$^{m}$58.524$^{s}$, and DEC=-00$^d$35$'$39.77$''$ with an uncertainty 
of 149 mas in RA and 114 mas in DEC.   
Considering the coordinates uncertainties, the positions of [C II] and 
dust continuum peaks are marginally consistent with that of the optical quasar (Figure 1). 
The quasar coordinates for SDSS 1044-0125 are RA=10$^{h}$44$^{m}$33.041$^{s}$, and DEC=-01$^d$25$'$02.09$''$, 
with an uncertainty of 27 mas in RA and 20 mas in DEC. This is well consistent with the positions of 
the [C II] and dust continuum peaks of this object (Figure 1). 

We finally check the results by comparing the measurements to those from the 
low-resolution ALMA Cycle 0 data (Table 1, \citealp{wang13}).
The [C II] line fluxes and continuum flux densities integrated over the source regions of both objects are 
consistent within the calibration uncertainties with the previous values. 
The ALMA data at $\rm 0.2''$ resolution reveal more detail on the spatial distribution and velocity field of 
the [CII]-emitting gas component in the quasar host galaxies. 

{\bf SDSS J0129-0035.} We present the intensity-weighted velocity map and velocity dispersion 
map of the [C II] line emission in Figure 3. 
We adopt the [C II] redshift of $\rm z_{[CII]}=5.7787$ as the zero velocity \citep{wang13}.
The line is single-peaked and the velocity map shows a velocity gradient from the southeast to 
northwest direction. The velocity images also show two features at the edge of the [C II]-emitting 
region (marked as No. 1 and 2 in the velocity map, Figure 3).  
One has an intensity-weighted velocity of $\rm -190\,km\,s^{-1}$ at the southeast edge of the 
line emitting region and the other has an intensity-weighted velocity of about $\rm +50\,km\,s^{-1}$ 
on the northwest part. There is no strong ($\rm \geq2\sigma$) dust emission detected in the regions of these two features. 
The intensity-weighted velocity dispersion map of
the [C II] emission (upper right panel of Figure 3) also suggests a high velocity dispersion structure along the
southeast-to-northwest direction, which seems to extend from the center to the two features. 
A structure with such a velocity dispersion cannot be explained with beam smearing of a
rotating disk. 

The PV diagrams along the direction of the two features 
and the perpendicular direction are plotted in the lower panel of Figure 3. 
The major gas component in the PV diagrams is consistent with rotation. 
The two features mentioned above are presented at the regions around (-0.4$''$, $\rm -190\,km\,s^{-1}$, No. 1 in Figure 3) 
and (+0.4$''$, $\rm +50\,km\,s^{-1}$, No. 2) on the bottom left panel. It is unclear if these two components are connected 
to the outer part of the rotating gas disk or not.  
The PV-diagrams also shows tentative peaks ($\rm 3\sim4\sigma$) with velocities of $\rm \sim\pm200\,km\,s^{-1}$ 
within the central 0.2$''$ (No. 3 and 4 in Figure 3). These components locate  
close to the center/peak of the [C II]-emitting region. However, the large velocities 
of $\rm \sim\pm200\,km\,s^{-1}$ seem not follow the velocity field the rotating component. 
These tentative peaks should be checked with deeper observations at better 
resolution (e.g., $\lesssim 0.1''$), to confirm if they are real features and related to gas infow/outflows.

We plot the [C II] line spectrum integrated over the line-emitting region in Figure 4, 
compared to the CO (6-5) line spectrum published in Wang et al. (2011). Though the CO (6-5) line profile 
is poorly constrained due to the low signal-to-noise ratio, the line widths and redshifts measured 
with the [C II] and CO lines are consistent with each other within the uncertainties \citep{wang11,wang13}.  
Moreover, comparing to the Gaussian line profile that fit to the line spectrum, 
there is a tentative excess on the blue wing around $\rm -200\,km\,s^{-1}$ (arrow in Figure 4). This is likely 
to be related to the components saw in the velocity map and PV diagrams (e.g., No. 1 and 3).

{\bf SDSS J1044-0125.} We calculate the intensity-weighted velocity and velocity dispersion maps with the $\rm >4\sigma$ 
pixels in each [C II] channel (Figure 5), which reveals the gas kinematics in the central high surface brightness part of the line-emitting 
region. The gas does show some velocity structure over this area. For example, the gas component 
to the northwest of the quasar position shows more positive velocities. However, the velocity map seems 
more complicated than what one would expect from a simple rotating disk. The PV-diagrams also suggest 
a very turbulent gas velocity field; the gas is distributed over a compact region of about 2.4 
kpc (i.e., $\rm \pm0.2''$) with velocities from -200 to $\rm \gtrsim200\,km\,s^{-1}$.  

The integrated [C II] line spectrum shows multiple peaks
and is not well fit by a single Gaussian profile (Figure 6). 
We try to fit the line profile with three Gaussian components (A, B, and C). The fitting results  
of the line centers, FWHM widths, and fluxes are listed in Table 2. Here, we adopt the [C II] redshift 
of $\rm z_{[CII]}=5.7847$ from the single Gaussian fitting as the zero velocity (Wang et al. 2013).
\citet{wang13} found a velocity offset between the Gaussian-fit line centers of the [C II] line 
and the CO (6-5) line for SDSS J1044-0125. With the new ALMA [C II] data in this work, we still see this offset. 
i.e., The CO (6-5) line overlaps only the blue part ($\rm -300-0\,km\,s^{-1}$) of 
the [C II] line emission with a line center at around $\rm -100\pm44\,km\,s^{-1}$ \citep{wang13} which is between the 
central velocities of components A and B. The mismatch between the [C II] and CO (6-5) spectra 
argues for complex gas components and kinematic properties in the nuclear region 
of SDSS J1044-0125 (see more discussion in Section 4.2).

On the velocity integrated map of the [C II] line (lower left of Figure 1 and left of Figure 7), 
there is a tentative peak $\rm 0.82''$ (i.e., 4.9 kpc) 
to the East of the major component. The spectrum of the emission from this secondary 
peak position shows positive signal from about -740 to $\rm 400\,km\,s^{-1}$. 
We integrate the emission in this velocity range and show the intensity map in the left panel of Figure 7. 
The secondary peak is detected at about 5$\sigma$ on the image, with a flux of $\rm 0.52\pm0.10\, Jy\,km\,s^{-1}$. 
We also check the Cycle 0 data and the image at $\rm 0''.6$ resolution does 
show an extension toward the secondary peak location (see red contours in the upper left panel of Figure 7).
A further check shows that this component is also 
present in a continuum-unsubtracted image in the same velocity range, as well as the image using only Cycle 1 data. 
No emission is detected at this location with the data from the other three continuum spectral windows.  
Fitting this component with a Gaussian profile yields a FWHM line width of $\rm 750\pm190\,km\,s^{-1}$,
centered at $\rm -190\pm80\,km\,s^{-1}$. The [C II] luminosity derived with the peak intensity of this component
is $\rm (4.8\pm0.9)\times10^{8}\,L_{\odot}$. 

\section{Analysis and Discussion}

\subsection{Surface brightnesses and ratios of the resolved [C II] and dust emission}

The ALMA [C II] images of SDSS J0129-0035 and SDSS J1044-0125 probe the ISM in the nuclear regions on kpc 
scales in these quasar-starburst systems. As we described in the previous section, 
the line and continuum emission are resolved from the two objects 
at $\rm 0.2''$ (i.e., $\rm \sim1.2$ kpc) resolution. The deconvolved 
FWHM source sizes of the [C II] line emission are 1.8$\pm$0.2 and 2.6$\pm$0.5 times larger than that of the continuum 
for SDSS J0129-0035 and SDSS J1044-0125, respectively (based on the results from imfit, Table 1). This is consistent with the results  
found in previous ALMA imaging of other starburst quasar host 
galaxies and SMGs \citep{wang13,diazsantos16,venemans16,venemans17c,gullberg18,cooke18,rybak19}. 
In most of these systems, the [C II] emission is found to be 1.3 to $>$2 times more extended than that of the 
dust continuum emission \citep{gullberg18,cooke18}.

Based on the [C II] and dust continuum intensity maps, we compare the surface brightnesses as a function of radius in the nuclear region of the 
two systems (Figure 8). We also include ULAS J1319+0950 in the comparison, which is another 
FIR luminous quasar-starburst system and spatially resolved in [C II] and dust continuum emission at a 
similar resolution in our ALMA Cycle 1 program (\citealp{shao17},see Section 1).  
Here we scale the continuum intensity map (Figure 1) with a factor of $\rm L_{FIR}/S_{con}$, 
where $\rm L_{FIR}$ is the FIR luminosity derived in \citet{wang13} 
assuming dust temperature of 47 K and emissivity index of $\rm \beta=1.6$ \citep{beelen06}, 
and $\rm S_{con}$ is the total continuum flux density close to the [C II] frequency in Table 1. 
This convert the monochromatic continuum map to a map of surface brightness of the FIR emission.
We calculate the mean [C II] and FIR surface brightnesses within the rings at central radii of 0$''$,
0.1$''$, 0.2$''$, 0.3$''$, 0.4$''$, 0.5$''$, 0.6$''$ and a ring width of 0.1$''$.
The scale of 0.1$''$ corresponds to a physical size of 0.58 kpc for ULAS J1319+0950, 
and 0.60 kpc for SDSS J0129$-$0035 and SDSS J1044$-$0125.
The symbols and error bars shown in Figure 8 represent the mean and standard 
deviation of the pixel values in each ring.
The [C II]-to-FIR surface brightness ratios are also plotted in the right panel of Figure 8. 
We model the surface brightness with an exponential disk in the 2-dimensional image 
plane with a S\'{e}rsic index of n=1. We then convolve the model with the synthesized beam, calculate 
the mean surface brightnesses of the model in each ring described above, and fit these mean 
values to the data. The models are shown as dashed lines in Figure 8. They suggest that the [C II] and dust emission in the 
nuclear region of the quasar host galaxies follow an exponential light profile. 
The resolved surface brightness ratios between the [C II] and FIR dust emission 
range from 0.0001 to 0.001 within the central region where both [C II] and 
dust continuum are detected. 

In Figure 9, we compare the resolved [C II] and FIR continuum emission at different radii in these three z$\sim$6 
quasars with other systems on the [C II]-FIR ratio vs. FIR surface brightness plot. We collect samples 
of star forming galaxies and quasars at low and high redshifts that have continuum source size measurements 
from ALMA, including the GOALS 
sample of local luminous infrared galaxies \citep{diazsantos13,diazsantos17,shangguan19}, the high-redshift 
SMGs \citep{riechers13,neri14,gullberg18}, Lyman Break 
Galaxies (LBG, \citealp{capak15,jones17,hashimoto18}), and the z$\sim$6 to 7 
quasars \citep{wang13,venemans16,venemans17c,diazsantos16,decarli18}. 
For the GOALS sample, we use the [C II]-FIR ratio extract from the central a few kpc 
region of the galaxies (i.e., the $\rm 9.4''\times9.4''$ region of one PACS detector 
element where the [C II] and continuum emission peaks, see details in \citealp{diazsantos13,diazsantos17}).
This is comparable to the [C II] and dust emission region detected in the high-z samples. 
We re-calculate the FIR luminosity surface brightness based on the GOALs sample
catalog\footnote{http://goals.ipac.caltech.edu/} and the decomposed infrared SED   
from \citet{shangguan19}. For the high-redshift samples that have a source size measurements from ALMA 
at $\rm 0.6''\sim0.7''$ resolution, we adopt the FIR surface 
brightness averaged within the half-light radius of the continuum emitting region. We also include the 
resolved [C II] and FIR emission from two submillimeter galaxies at z=2.9, which were imaged with ALMA at $0.16''$ 
resolution and have FIR luminosities and source sizes comparable to the two quasar host galaxies we studied 
in this work \citep{rybak19}. 

The continuum emission from SDSS J0129$-$0035 and SDSS J1044$-$0125, as well as ULAS 
J1319+0950, was resolved by ALMA on 1$\sim$2 kpc scale and the peak FIR surface brightness 
is up to $\rm \Sigma_{FIR}\sim2\times10^{12}\,L_{\odot}\,kpc^{-2}$. 
This is comparable to the highest values found in the compact SMGs and ULIRGs. 
As was shown in Figure 9, at this FIR surface brightness, the [C II]-FIR ratio is typically 
an order of magnitude lower than that of normal star-forming galaxies.
The [C II]-FIR ratios in the three quasar hosts get higher toward larger radii, and the FIR surface 
brightnesses decrease to a few $\rm 10^{10}\,L_{\odot}\,kpc^{-2}$. 
Thus the radial change of the [C II]-FIR ratios found with the 
three objects can be understand as a trend of descreasing [C II]-FIR ratio  
with increasing FIR surface brightness which was found in samples of 
IR luminous star forming systems (e.g., $\rm \Sigma_{FIR}\,\geq$ 
a few $\rm 10^{10}\,L_{\odot}\,kpc^{-2}$, \citealp{herreracamus18}). 
This trend have been termed "[C II]-FIR deficit" 
and was discussed in a number of papers \citep{stacey10,haileydunsheath10,ivison10,farrah13,gullberg15,gullberg18,diazsantos16,
lutz16,smith17,herreracamus18,cooke18}. 

In the compact dusty starburst systems, the high FIR surface 
brightness reveals high star formation rate surface density with high gas density and strong UV radiation field. 
As was mentioned in Section 1, in such environment, 
the [C II] emission could be suppressed due to several possible mechanisms, 
including e.g., saturation of the [C II] transition 
when the gas is heated with temperatures above 91 K, \citep{munoz16,rybak19}, 
increase of dust absorption of UV photoms with high ionization
parameter and a decrease of photoelectric heating efficiency \citep{herreracamus18}, or   
reduced amount of C$^+$ medium with more carbon in form of CO at high cloud surface 
density \citep{narayanan17}. 
Though a model study of the origin of the [C II]-FIR deficit is beyound the goals of this work, 
we can conclude that the low [C II]-FIR ratio and high FIR surface brightness found in the central region of
the FIR luminous quasar hosts suggest a similar high density and dusty ISM heated by a high-intensity 
radiation field as was found in the compact and intense starburst systems. 

Of the three objects, SDSS J1044$-$0125 and ULAS J1319+0950 are about 10 times
brighter than SDSS J0129-0035 in quasar UV luminosity \citep{jiang09,shen19}. 
However, their [C II]/FIR surface brightnesses and luminosity ratios are comparable. 
Though the central energetic AGN with a hard UV radiation field and 
strong X-ray emission may contribute to the ISM ionization and heating in the 
nuclear region \citep{stacey10,smith17,shangguan18,herreracamus18}, it is unlikely 
to be the dominant cause of the [C II]-FIR deficit in the quasar host galaxies (see also \citealp{decarli18}).  
As was pointed out by the spatially resolved studies of SMGs \citep{gullberg18, rybak19} and local AGNs \citep{smith17}, 
the change of [C II]/FIR ratio with FIR surface brightness at different radii in the quasar host galaxies suggest that 
the [C II]/FIR deficit depends on the local conditions (radiation field, gas density,etc.), 
rather than the global properties of the systems.

We did not detect dust continuum emission at the [C II] peak to the east of the quasar SDSS J1044-0125.
The continuum image sets a 3$\sigma$ upper limit of 0.12 mJy for the flux density at the [C II] frequency.
Assuming dust temperatures of 25 to 45 K, this constrains
the FIR luminosity of this component to be $\rm <2\times10^{11}\, L_{\odot}$, and a [C II]-to-FIR luminosity
ratio of $\rm > 2.4\times10^{-3}$ which is much higher than the central region of the quasar host.
The high [C II]-to-FIR ratio of this component reveals ISM conditions different from that in
the intense starburst nuclear region of the quasar host galaxy. Deep imaging of molecular CO and other
fine structure line emission will help to measure the physical properties and excitation of the medium at this offset [C II] peak
and address the nature of this component (i.e., core of mergering galaxy or node of outflow, etc. see more discussions below).

\subsection{Gas kinematics in the nuclear region}

The [C II]-emitting gas in SDSS J0129-0035 shows clear velocity
gradients, suggesting that most of the gas is likely from a rotating disk. The ordered motion
constrains the dynamical mass of the quasar host galaxy within the [C II]-emitting region (see details Section 4.3).
The $\rm 0.2''$ resolution [C II] image of SDSS J0129-0035 also reveals additional gas clumps, in particular,
the components in the $\leq0.2''$ region of the PV-diagram with velocities
of $\rm \sim\pm200\,km\,s^{-1}$ (Marked as No. 3 and 4 in the lower panel of Figure 3)
and the two extended features as were described in Section 3. 
Though the individual tentative peaks with limited ($\rm 3\sim4\sigma$, Figure 3) 
SNR should be checked with deeper observations, we argue that the rotating velocity field 
with turbulent clumps reveals complex kinematics of the gas component in the nuclear region. 
Note that similar turbulent rotating gas disks were also reported 
with the CO(6-5) and [C II] images of SDSS J231038.88+185519.7 at z=6.0 \citep{feruglio18} 
and the [C II] image of VIKING J030516.92$-$315056.0 at z=6.6 \citep{venemans19} 
and was discussed in recent numerical simulations (e.g., \citealp{lupi19}); the gas around  
these powerful quasars show significant dynamical evolution with sub-structures, including 
spirals, outflow and inflow clumps, and/or truncated disk.

The [C II] map of SDSS J1044-0125 does not show a clear sign of rotation. The PV-diagrams (Figure 5) suggest a very
turbulent gas velocity field in the central area; the gas within the $\rm 0.2''$ region shows velocities
ranging from -200 to $\rm \gtrsim200\,km\,s^{-1}$. In addition, the [C II] line spectrum suggest multiple peaks (Table 2). 
Comparison between the CO (6-5) and [C II]158$\mu$m line
spectra of this object (Figure 6) shows that the CO (6-5) line does not
cover the same velocity range as that of the [C II] line. The high-J CO (e.g. CO (6-5), (7-6)) and [C II] lines
in other millimeter-bright quasars at z$\sim$6 usually show similar velocities and line
widths \citep{riechers09,wang13}, suggesting that these lines trace similar kinematics of
the ISM from the nuclear star forming region. The case of SDSS J1044-0125 suggest a multiple-phase ISM with different
kinematic properties. The [C II] emission in galaxies is reported to trace the disk gas and star forming regions, as well
as the warm diffuese neutral/ionized medium \citep{stacey10,pineda14,gullberg15,croxall17,decarli18,lagache18}.
[C II] emission in the quasar host galaxies could also
powered by the central powerful AGN (e.g., from the narrow line
region, the X-ray dominated regions, etc \citep{stacey10,uzgil16}).
Asymmetric [C II] line emission that is broader than the molecular CO emission was detected the
radio galaxy 3C 326N, suggest [C II] emission from a warm diffuse and turbulent molecular gas
component powered by AGN-jet activity \citep{guillard15}. On the other hand, the CO(6-5) line emission found in
the high-z quasar-starburst systems usually traces the dense molecular
gas where stars are actively forming \citep{greve14,liu15,riechers09,venemans17a}.
Thus, it is possible that, in SDSS J1044-0125, most of the [C II] emission covering a similar velocity
range with the CO (6-5) line is from the intense star forming regions, while the additional components
in the velocity range without CO detection may be from more diffuse/warm medium.
This mismatch between the CO and [C II] line spectra is a curious feature and should be
checked with deeper and high-resolution imaging of the molecular line emission as well
as other fine structure lines to address the complex physical condition, power resource,
and kinematics of the ISM in this region.

The peak to the east of SDSS J1044-0125 reveals a [C II]-emitting source with a very broad line
width of $\rm FWHM=750\pm190\,km\,s^{-1}$ at a projected distance of 4.9 kpc from the quasar (Component II in Figure 7).
The nature of this offset component is unclear yet. Broad (line width up to $\rm \geq1000\,km/s$),
blue/red-shifted, and low surface brightness [C II] emission was detected as gas outflows in the
young quasar hosts at high-redshift, which are evidences of strong AGN feedback \citep{cicone14,cicone15,fiore17,bischetti18}.
The offset peak we found in SDSS J1044-0125 could be a bright node of such outflowing gas component.
Companion [C II]/submm continuum sources
are detected close to other luminous quasars at high redshift \citep{decarli17,trakhtenbrot17},
revealing that these massive objects are in the centers of very active environments and that galaxy mergers may be a major
triggering mechanism for SMBH and galaxy growth in these systems \citep{li07,narayanan08,narayanan15,
volonteri15,watson15,valiante17,marrone18}. SDSS J1044-0125, as one of
the most luminous quasars known at this redshift, may lie in a massive dark matter halo,
and has already experienced several galaxy merger events \citep{li07,valiante17,marrone18}.
Thus, it is also possible that this offset [C II] component is the dense core of a satellite
galaxy that will merge with the quasar host. 

In summary, the ALMA observations at 0.2$''$ resolution reveal complex gas kinematics 
in the nuclear region of the quasar starburst systems at z$\sim$6, e.g., rotation and/or turbulent gas clumps. 
The turbulent gas kninematis, as well as the multiphase ISM suggested by the [C II] and CO emission 
in SDSS J1044-0125, indicate rich dynamical activities of ISM powered by the central AGN and/or the intense 
nuclear star formation. It will be interesting to expand the high resolution imaging 
studies to more examples to investigate the kinematic activities of the ISM in these extreme 
quasar-starburst systems. The nature of the peak to the east of SDSS J1044-0125 should also be checked with deeper 
observations of other emission lines to constrain the ISM properties and its dynamical connection to the quasar host.

\subsection{Constraints on the host galaxy dynamics of SDSS J0129-0035}

We here model the gas velocity map of SDSS J0129-0035 with a simple
circular rotating gas disk, following the method used in \citet{shao17} (see also \citealp{jones17}), to constrain the
gas dynamics in this system. We use the ROTCUR task in the Groningen Image Processing 
System (GIPSY3; \citealp{vanderhulst92}) to fit the observed velocity field. As was described in \citet{shao17},
the parameters for the model are: (i) The sky coordinates of the rotational center of the disk. 
(ii) The system velocity $\rm V_{sys}$. (iii) The circular velocity at the fitting radius $\rm V_c(R)$. (iv) 
The position angle of the major axis on the receding half of the disk, anti-clockwise from the 
north direction, $\rm \phi$. (v) The inclination angle between the normal direction 
of the ring and line of sight, {\em incl}, and (vi) the azimuthal angle related to $\rm incl$, 
$\rm \phi$, and the disk center. We fix the disk center as the Gaussian peak center of the [C II] emission, 
and choose $\rm V_{sys}=0\,km\,s^{-1}$ to refer to the [C II] redshift of 5.7787 \citep{wang13}. We then use the data within a 
radius of 0.35$''$ to determine $\rm \phi $ and {\em incl}. This is approximately the $\rm 2\sigma$ contour 
region of the [C II] line emission (Figure 1). This fit gives $\rm \phi $ and {\em incl} as 
$\rm \phi=56\pm1^{\circ}$ and $\rm incl=16\pm20^{\circ}$, respectively, where the error bars denote the fitting errors.  
This suggests a gas disk that is very close to face-on with incl$\leq36^{\circ}$. 

In \citet{wang13}, we adopted the same inclined circular disk assumption and estimated the inclination 
angle from the deconvolved axis ratio of the Cycle 0 [C II] map, i.e., 
$\rm incl=arccos{(a_{minor}/a_{major})}=56^{\circ}$. This is much larger than the value we obtained 
in this paper. It is likely that the two extended features (No. 1 and 2 in Figure 3) 
are unresolved in the low-resolution ALMA Cycle 0 image, which results 
in a larger major axis value, i.e., $\rm a_{major}=0.41''\pm0.06''$ from the Cycle 0 data compared 
to $\rm a_{major}=0.32''\pm0.03''$ in this work (Table 1). We cannot determine
if these two features are part of the rotating disk or not, thus, we did not include those 
pixels in the fitting in this work.

Adopting the $\rm \phi $ and {\em incl} values above, we constrain the circular velocities at 
radii of 0.1$''$, 0.2$''$, and 0.3$''$ (Figure 10), corresponding to physical scales of 0.6 kpc, 1.2 kpc, and 1.8 kpc. 
If the inclination angle is fixed to 16$^{\circ}$, the circular velocity at the largest radius (i.e., 1.8 kpc) 
 is $\rm V_c$=250 $\rm km\,s^{-1}$. Considering the fitting uncertainties of {\em incl}, we estimate a lower limit 
to the circular velocity of $\rm V_c$=110 $\rm km\,s^{-1}$ with $\rm incl=36^{\circ}$. An upper limit of the circular velocity cannot be 
constrained as {\em incl} could be very close to zero. Using $\rm V_c=250\, km\,s^{-1}$, 
we derive the dynamical mass within a radius of 1.8 kpc to be $\rm M_{dyn}=2.6\times10^{10}\,M_{\odot}$. 
If the lower limit of the circular velocity of 110 $\rm km\,s^{-1}$ is adopted, the dynamical mass 
will descrease to $\rm 5.2\times10^{9}\,M_{\odot}$. 

Based on the quasar 1450$\AA$ magnitude from the SDSS photometry \citep{jiang09} and a 1450$\AA$ 
bolometric correction of $\rm L_{bol}=4.2\times\lambda L_{\lambda1450}$ \citep{runnoe12}, we detrive a bolometric 
lumiosity of $\rm5.7\times10^{12}\,L_{\odot}$ for SDSS J0129-0035. 
Assuming that the SMBH is accreting at the Eddington limit, we estimate the SMBH mass of this 
object to be $\rm M_{BH}=1.7^{+1.7}_{-0.9}\times10^{8}\,M_{\odot}$. 
We consider an uncertainty of 0.3 dex in the SMBH mass, which represents the scatter of the Eddington 
ratio of the z$\sim$6 quasar sample \citep{derosa11}.  
This result in a SMBH to host galaxy 
dynamical mass ratio of $\rm M_{BH}/M_{dyn}=0.007$ with $\rm V_c=250\, km\,s^{-1}$. If the lower limit of $\rm V_c=110\,km\,s^{-1}$ is adopted, 
the mass ratio will become 0.03.  
Considering the facts that i) the model has large uncertainties for face-on system and cannot accurately determine 
the inclination angle and circular velocity, ii) we do not reach the flat part 
of the rotation curve as shown in Figure 10, and iii) the outer part of the disk may not be traced by the [C II] velocity 
map due to low signal-to-noise ratio or is disrupted \citep{venemans19,lupi19}, the $\rm M_{dyn}$ value derived 
above should be considered as a lower limit of the host galaxy dynamical mass for SDSS J0129-0035. 
The corresponding $\rm M_{BH}/M_{dyn}$ ratio hence represents an upper limit.  

We compare the SMBH mass and the lower limit of the host galaxy dynamical mass of SDSS J0129-0035 to those of other 
quasars at z$\sim$6 to 7 quasars and the relationship of local galaxies \citep{kormendy13}.
We collect the dynamical mass measurements of the z$\sim$6 to 7 quasars from the literatures. 
These objects are all detected in [C II] and/or CO line emission \citep{wang13,wang16,cicone15,willott15,willott17,venemans16, 
feruglio18,izumi18,decarli18,wang19} and have source size measurements 
from [C II] or CO data and inclination angles estimated from the minor and major axis ratios of the line emission. 
The two objects, SDSS J0129-0035 (this work) and ULAS J1319+0950 \citep{shao17}, were observed with ALMA 
at a much better resolution of $\rm 0.2''\sim0.3''$. The [C II] emission is resolved and the velocity maps 
show clear velocity gradients. These, combined with the disk dynamical model described above, 
provide more reliable constraints on the disk inclination angle, circular velocity, 
and dynamical mass of the quasar host galaxies. We denote these two objects as blue 
and red stars in the plot. According to Figure 11, the black hole-to-host mass relationship 
of local galaxies given by \citet{kormendy13} predicts a  
mass ratio of 0.0043 for a SMBH mass of $\rm 1.7\times10^{8}\,M_{\odot}$. 
Thus, it is unlikely that the black hole-to-host mass ratio of SDSS J0129-0035 is significantly 
above the local relationship. This is unlike the most luminous/massive quasars at this redshift, which usually 
show ratios a few to $\gtrsim10$ times higher \citep{venemans16,decarli18,wang19}. 
This agrees with the results of other [C II]-detected z$\gtrsim$6 quasars with similar or lower SMBH 
masses, in which the ratios are consistent with the trend defined by the 
local systems \citep{willott10,willott17,izumi18}. 
The dynamical mass constraints based on the [C II] or CO imaging of these z$\sim$6 to 7 quasars suggest that, in the early universe, 
the most massive SMBHs with masses of $\rm 10^{9}$ to $\rm 10^{10}\,M_{\odot}$ may grows 
faster than that of their host galaxies \citep{walter04,venemans16,wang16,decarli18,wang19} 
while the less massive systems ($\rm 10^7\sim10^{8}\,M_{\odot}$) are evolving more closer to 
the trend of local galaxies \citep{willott17,izumi18}.

\section{Summary}

Based on new ALMA observations we present images of the [C II] and dust continuum emission from the host galaxies of 
two z=5.78 quasars, SDSS J0129$-$0035 and SDSS J1044$-$0125 on $\rm 0.2''$ scales, resolving the ISM emission in the 
central $\rm \sim4\,kpc$ of the hosts. The main conclusions are summarized as follows: 

\begin{itemize}
\item  The FWHM sizes of the [C II] emission is 1.8-2.4 kpc and that of the 
dust continuum $\sim$1kpc for the two bojects, constraining the size 
of the intense star-forming region around the accreting SMBHs. As was found in other 
starburst quasar host galaxies and SMGs at high redshift, 
The [C II] emission regions appear to be more extended compared to that of dust continuum \citep{gullberg18,cooke18}.  
\item We derive the [C II] and FIR surface brightness and emission ratios at different 
radii based on the resolved [C II] and continuum images. 
The resolved [C II]-FIR ratios are decreasing with increasing $\rm \Sigma_{FIR}$ toward the center, 
following the trend of the IR luminous star forming galaxies with $\rm \Sigma_{FIR}$ 
in the same range of a few $\rm 10^{10}\,L_{\odot}\,kpc^{-2}$ 
to $\rm 2\times10^{12}\,L_{\odot}\,kpc^{-2}$ \citep{stacey10,gullberg15,gullberg18,
diazsantos13,lutz16,herreracamus18,cooke18,rybak19}. The low [C II]-FIR ratios of $0.0001\sim0.0002$ found 
at the peak $\rm \Sigma_{FIR}$ in the central region of these two quasar hosts  
suggest dense ISM with high-intensity radiation field similar to that in 
the compact dusty starburst systems \citep{diazsantos17,herreracamus18,gullberg18,rybak19}.  

\item The velocity field and PV diagrams of SDSS J0129-0035 reveal a combination 
of rotation with turbulent gas clumps; We detect clear velocity gradients in the velocity map. 
We also see features with velocities offsets of -190 and $\rm +50\,km\,s^{-1}$ at the 
outer part of the [C II] emitting region and gas clumps with velocity offsets 
of $\rm \pm200\,km\,s^{-1}$ within the central 0.2$''$ region, suggesting 
other kinematic activities, such as inflow or outflows.

\item The [C II]-emitting gas in the central region of SDSS J1044-0125 does not show 
a clear sign of rotation. The PV-diagram of the [C II] emission 
suggests very turbulent gas kinematics in the central area. The [C II] line spectrum shows multiple peaks, which is 
broader and slightly offset from the CO (6-5) line emission. 
In addition, we tentatively detect an offset peak, 4.9 kpc
away to the east of the quasar with a very broad line width ($\rm 750\pm190\,km\,s^{-1}$) which may relate 
to an infalling companion source or node of outflows.

\item Overall, the gas kinematics on kpc scale in the nuclear region 
of the two starburst quasar host galaxies reveal significant dynamical 
evolution of the ISM, which could be powered by both AGN and the intense star formation.

\item Based on the dynamical modeling of the [C II] velocity gradients of SDSS J0129$-$0035, 
we derive a disk inclination angle very close to face-on
(i.e., $\rm incl=16\pm20^{\circ}$). We can thus only put a lower limit on the host dynamical mass
within the [C II]-emitting region. The result argues against an SMBH-bulge mass ratio that is significantly
above (e.g., larger by a factor of 10) the value derived from the local relationship. 
As was discussed in previous studies of the less luminous/massive quasar population at z$\sim$6, 
the mass ratio of SDSS J0129-0035 is likely to be closer to the local
relationship than to that found for the most luminous and massive quasars at the same
redshift \citep{venemans16,willott10,izumi18,wang19}.
\end{itemize}

The ALMA images of the [C II] emission in these young quasar host galaxies at kpc resolution 
reveal rich features of gas kinematics in the nuclear region around the powerful AGNs. 
Between the two objects studied in this work, we do see more turbulent/clumpy gas in the 
nuclear region and an curious discrepancy between the [C II] and CO line profiles in SDSS J1044$-$0125, 
which hosts a more massive SMBH with higher quasar luminosity. 
However, it is still unclear if this is due to the feedback of the powerful AGN or nuclear star formation. 
If the east peak is associated with infalling companion, the tidal perturbations could also 
affect the gas dynamic in the nuclear region. 
Deeper images of the [C II] and molecular CO emission from this object are still required to confirm these features.   
The high-resolution imaging studies should also be expanded to a larger sample to 
check for significant evolution of gas kinematic properties with AGN activity, e.g., more turbulent or disrupting gas disk 
in more luminous AGNs. Deep imaging of the 
molecular gas (e.g., with ALMA) and the stellar component (e.g., with the JWST in the future) on similar scales are also 
urgently required to fully probe host galaxy evolution of these young quasars at the earliest epochs.   

\acknowledgments

This work was supported by National Key Program for Science and Technology
Research and Development (grant 2016YFA0400703). We are thankful for the supports from the National
Science Foundation of China (NSFC) grants No.11721303, 11373008, 11533001,
the Strategic Priority Research Program $'$The Emergence of Cosmological Structures$'$
of the Chinese Academy of Sciences, grant No. XDB09000000,
the National Key Basic Research Program of China 2014CB845700. RW acknowledge supports from the Thousand Youth
Talents Program of China, the NSFC grants No. 11443002 and 11473004
D.R. acknowledges support from the National Science Foundation under grant number AST-1614213.
DN was funded in part by NSF AST-1715206 and HST AR-15043.0001.
The National Radio Astronomy Observatory (NRAO) is a facility of the National
Science Foundation operated under cooperative agreement by Associated
Universities, Inc. This paper makes use of the following ALMA data:
ADS/JAO.ALMA\#2011.0.00206.S and \#2012.1.00240.S . ALMA is a partnership
of ESO (representing its member states), NSF (USA) and NINS (Japan),
together with NRC (Canada) and NSC and ASIAA (Taiwan), in
cooperation with the Republic of Chile. The Joint ALMA Observatory
is operated by ESO, AUI/NRAO and NAOJ. RW thanks Dr. Matus Rybak 
for providing the resolved [C II] and FIR continuum measurements 
of their SMG sample. RW thanks Dr. Shangguan for providing FIR luminosities of the GOALS sample.

\email{rwangkiaa@pku.edu.cn}.

{\it Facilities:} \facility{ALMA}.

\clearpage

\begin{figure}
\plottwo{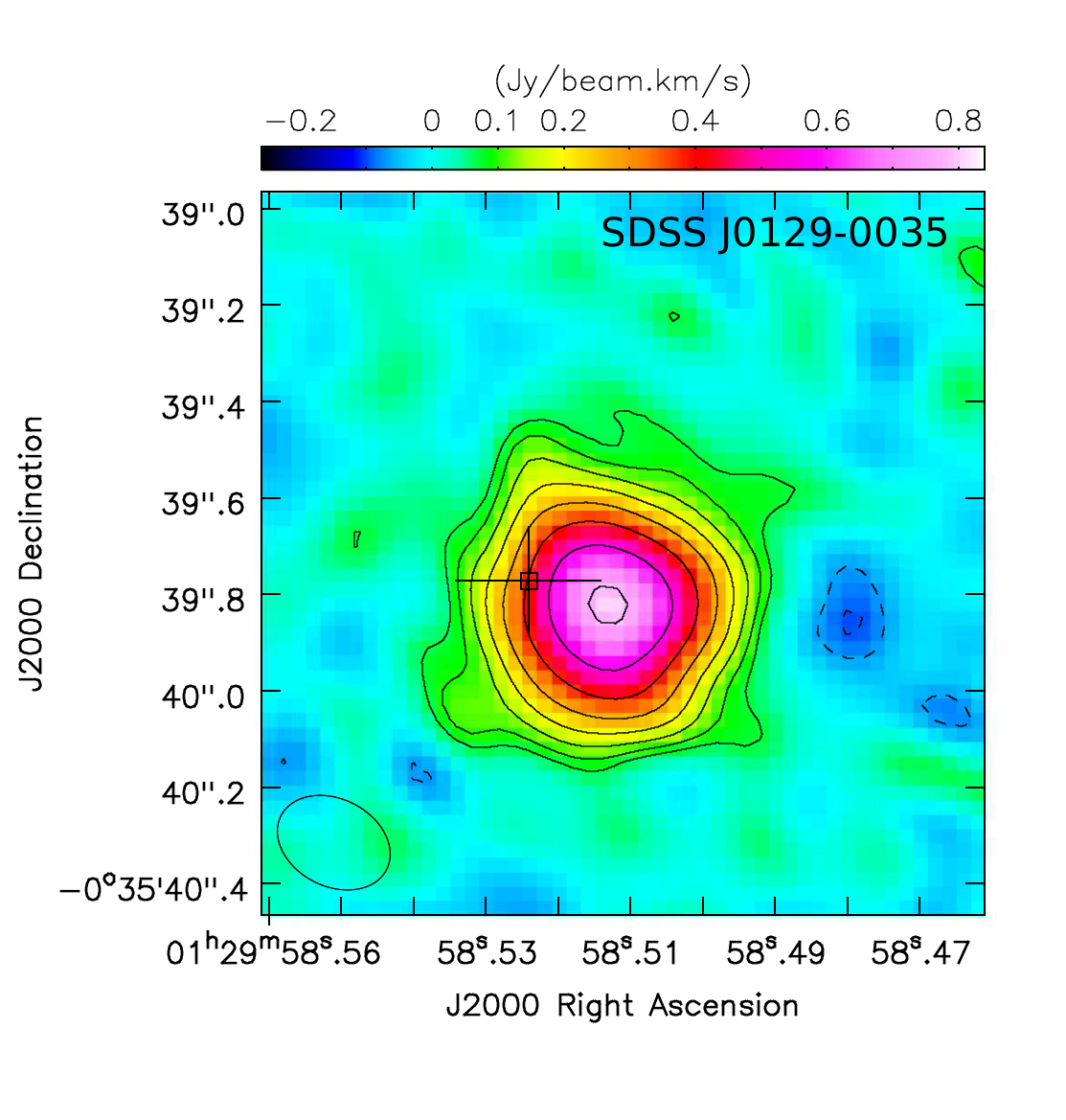}{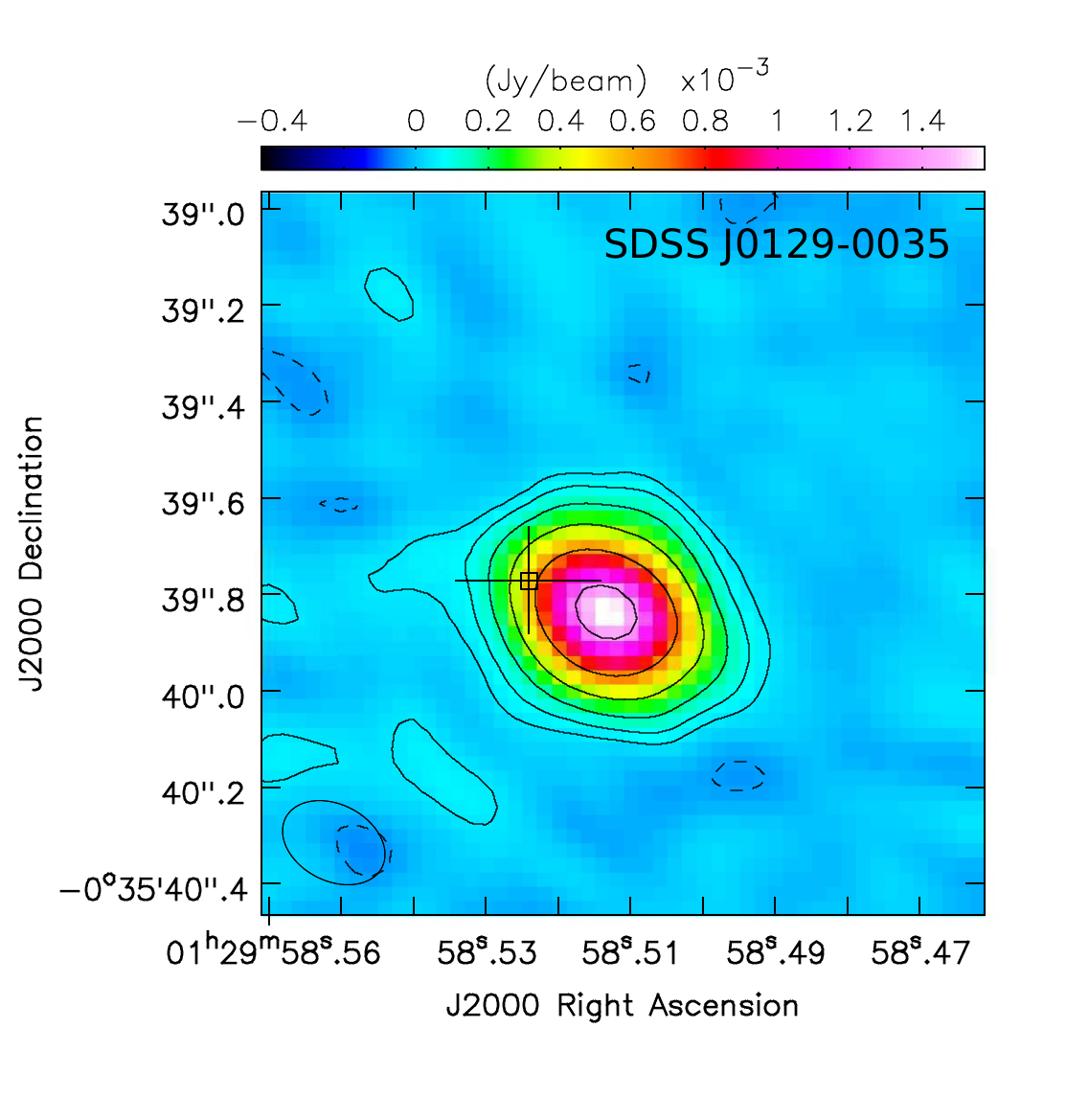}\\
\plottwo{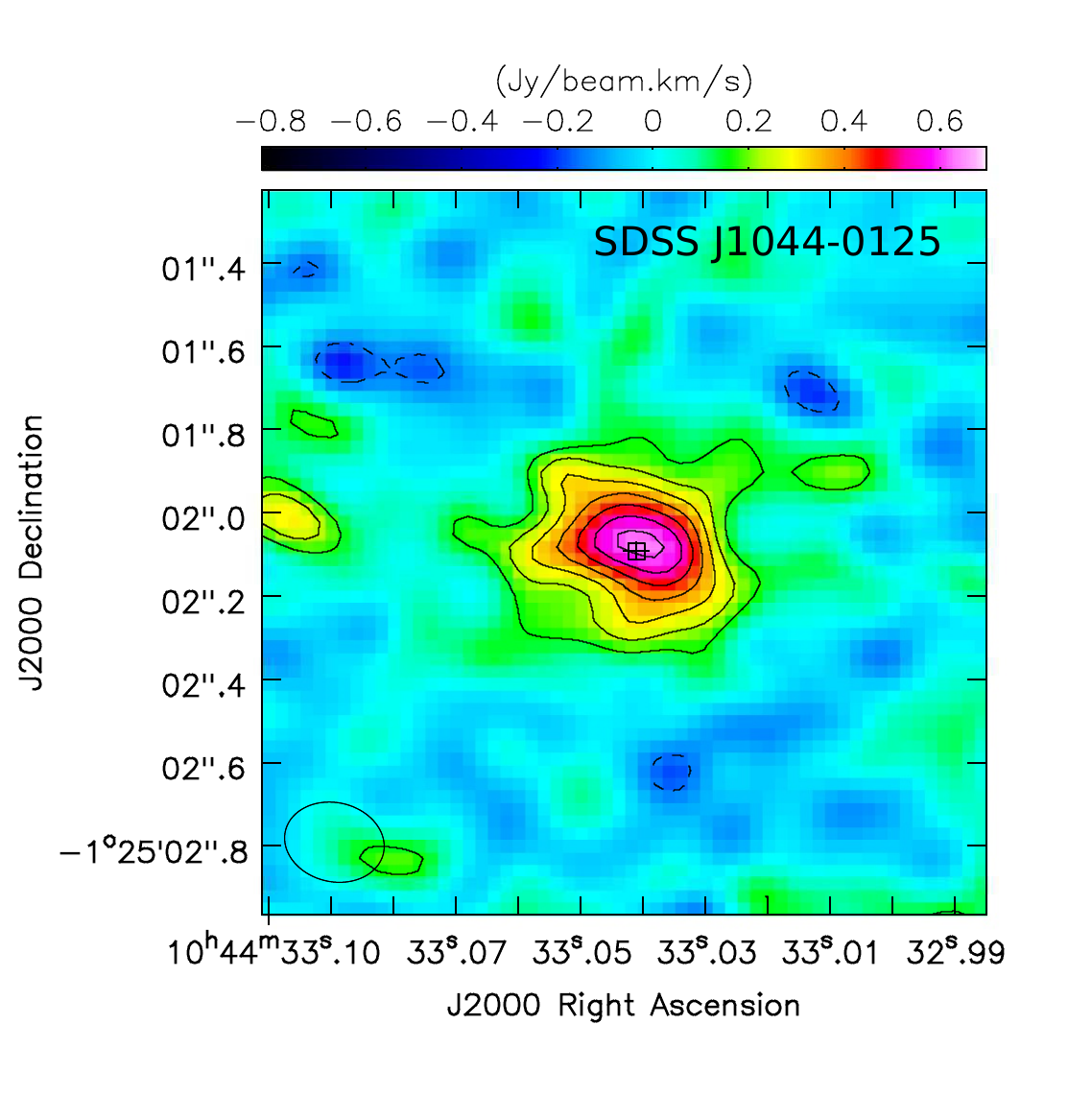}{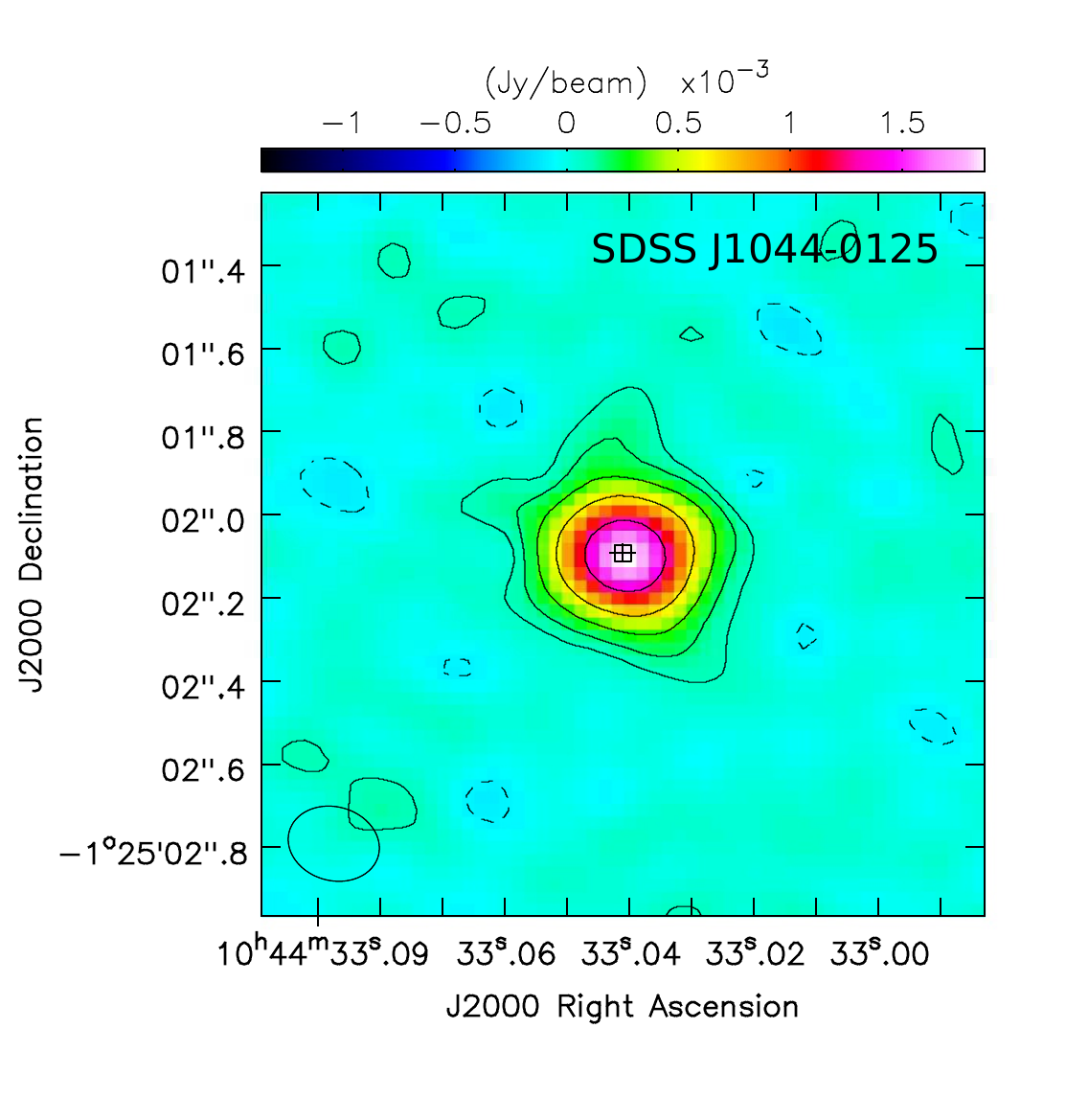}
\caption{The upper panels show the [C II] line velocity-integrated map (left) and continuum intensity map (right)  
of SDSS J0129$-$0035. The contours are [-2$\sqrt{2}$, -2, 2, 2$\sqrt{2}$, 
4, 4$\sqrt{2}$, 8, 8$\sqrt{2}$, 16, 16$\sqrt{2}$]$\rm \times0.034\,Jy\,beam^{-1}\,km\,s^{-1}$ for the line, 
and [-2, 2, 4, 8, 16, 32, 64]$\rm \times0.021\,mJy\,beam^{-1}$ for the continuum. 
The ellipse on the bottom left denotes the FWHM synthesized beam size, which 
is $\rm 0''.25\times0''.18$ for the line and $\rm 0''.22\times0''.16$ for the continuum. 
The black squre with a cross represents the position of the optical quasar, and the cross bars 
represent the uncertanties of 149 mas in RA and 114 mas in DEC measured from GAIA.
The lower panels show the [C II] line velocity-integrated intensity map (left) and continuum map (right) of SDSS J1044$-$0125.
The contours are [-2, 2, 3, 4, 5, 6, 7, 8]$\rm \times0.076\,Jy\,beam^{-1}\,km\,s^{-1}$ for the line, 
and $\rm [-2, 2, 4, 8, 16, 32]\times0.04\,mJy\,beam^{-1}$ for the continuum. 
The FWHM beam sizes shown on the bottom left are $\rm 0''.24\times0''.19$ for 
the line and $\rm 0''.22\times0''.18$ for the continuum.
The black squre with a cross denote the optical quasar position and the cross bars
represent the uncertanties of 27 mas in RA and 20 mas in DEC from GAIA.}
\end{figure}

\begin{figure}
\plottwo{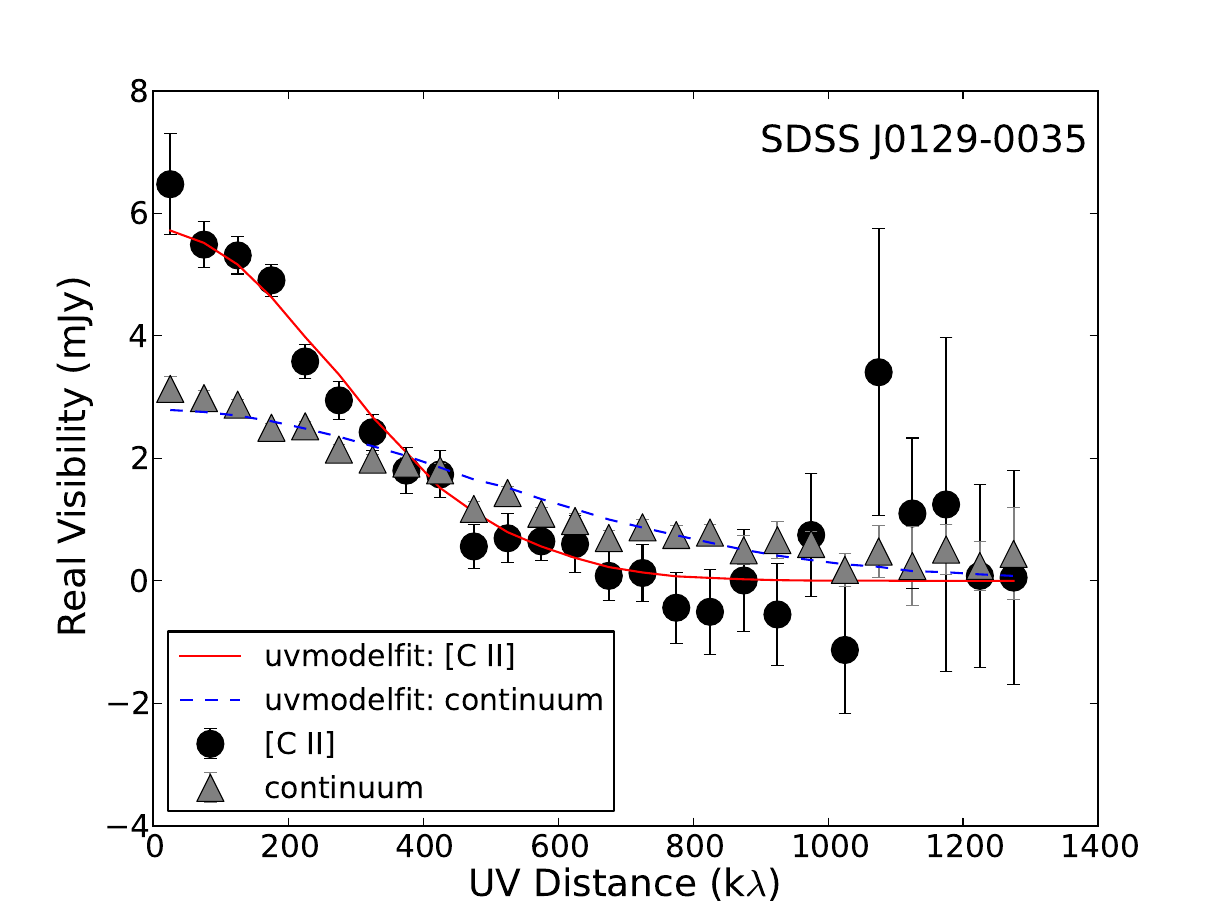}{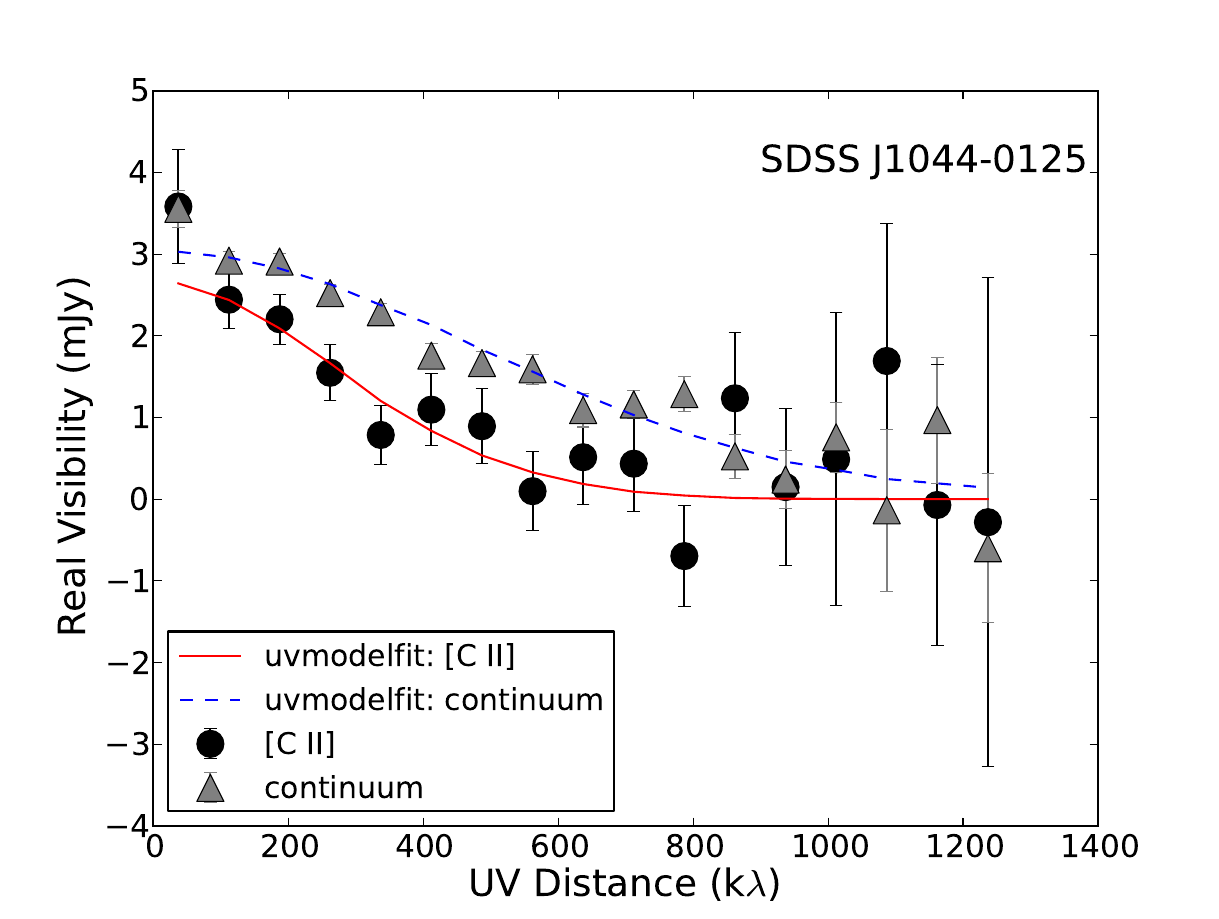}
\caption{Visibility versus uv distance of the [C II] line (black squares) and continuum 
emission (gray triangles). We re-set the phase center to the peak of the [C II] line emission (Table 1).  
The visibility data are radially binned with a bin size of 
50 k$\lambda$ for SDSS J0129-0035 and 75 k$\lambda$ for SDSS J1044-0125. 
The error bars show the standard deviation of the mean value in each bin. 
We fit an elliptical Gaussian to the uv data and the uv models are shown as red solid and blue 
dashed lines for the line and continuum, respectively.  
}
\end{figure}

\begin{figure}
\includegraphics[height=2.8in]{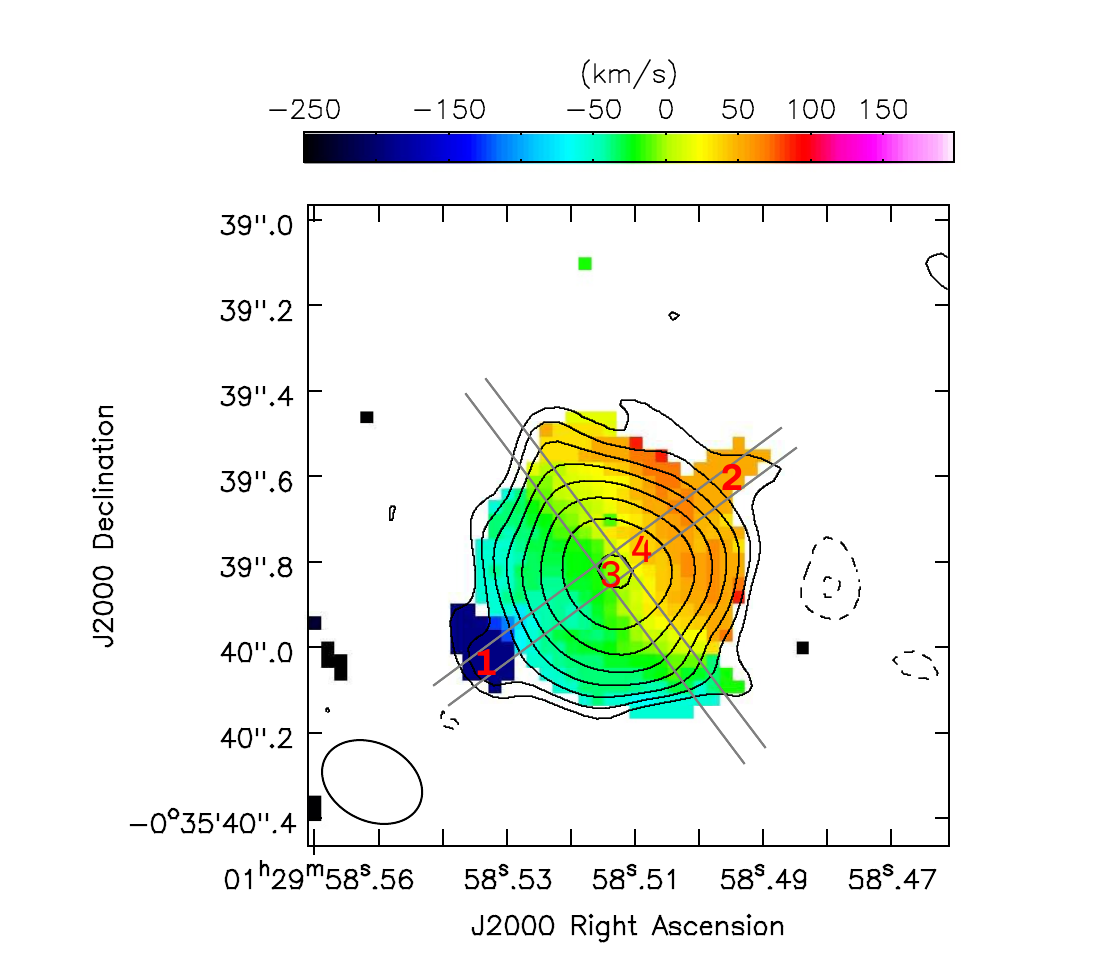}
\vskip -2.8in
\hspace*{3.0in}
\includegraphics[height=2.8in]{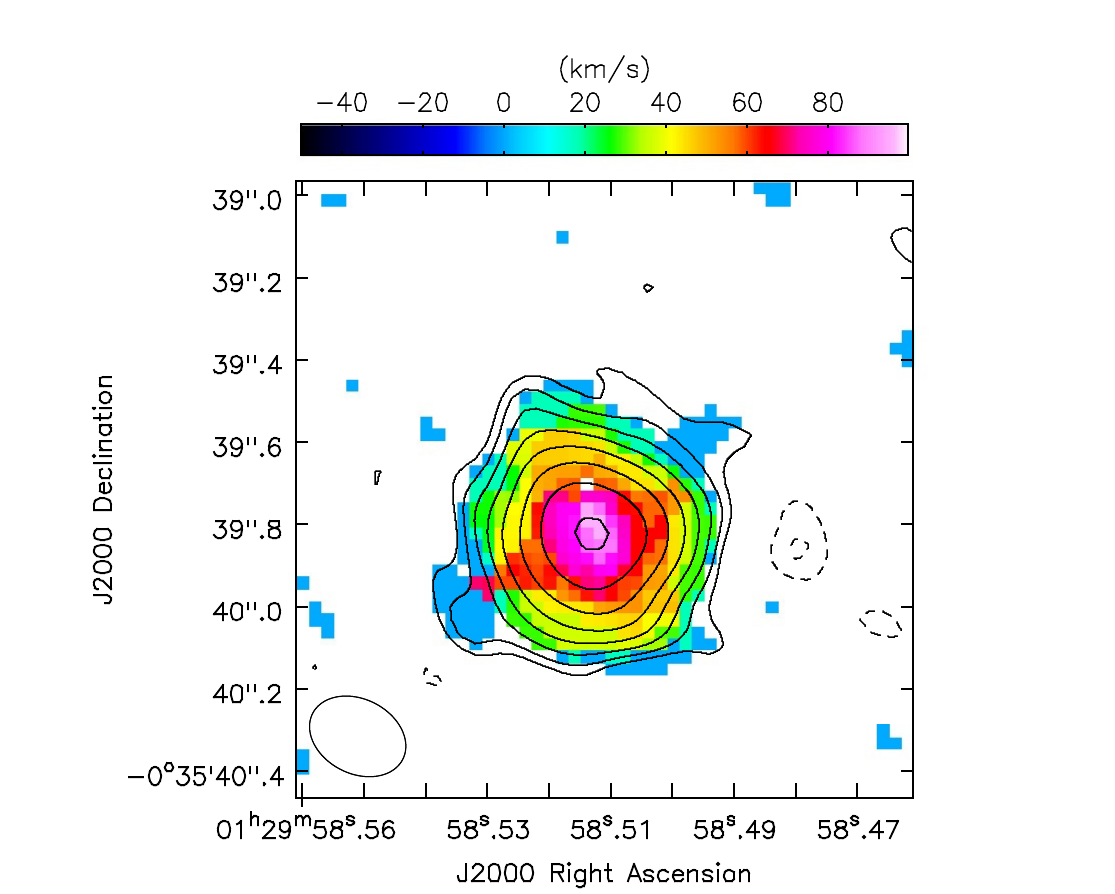}\\
\plottwo{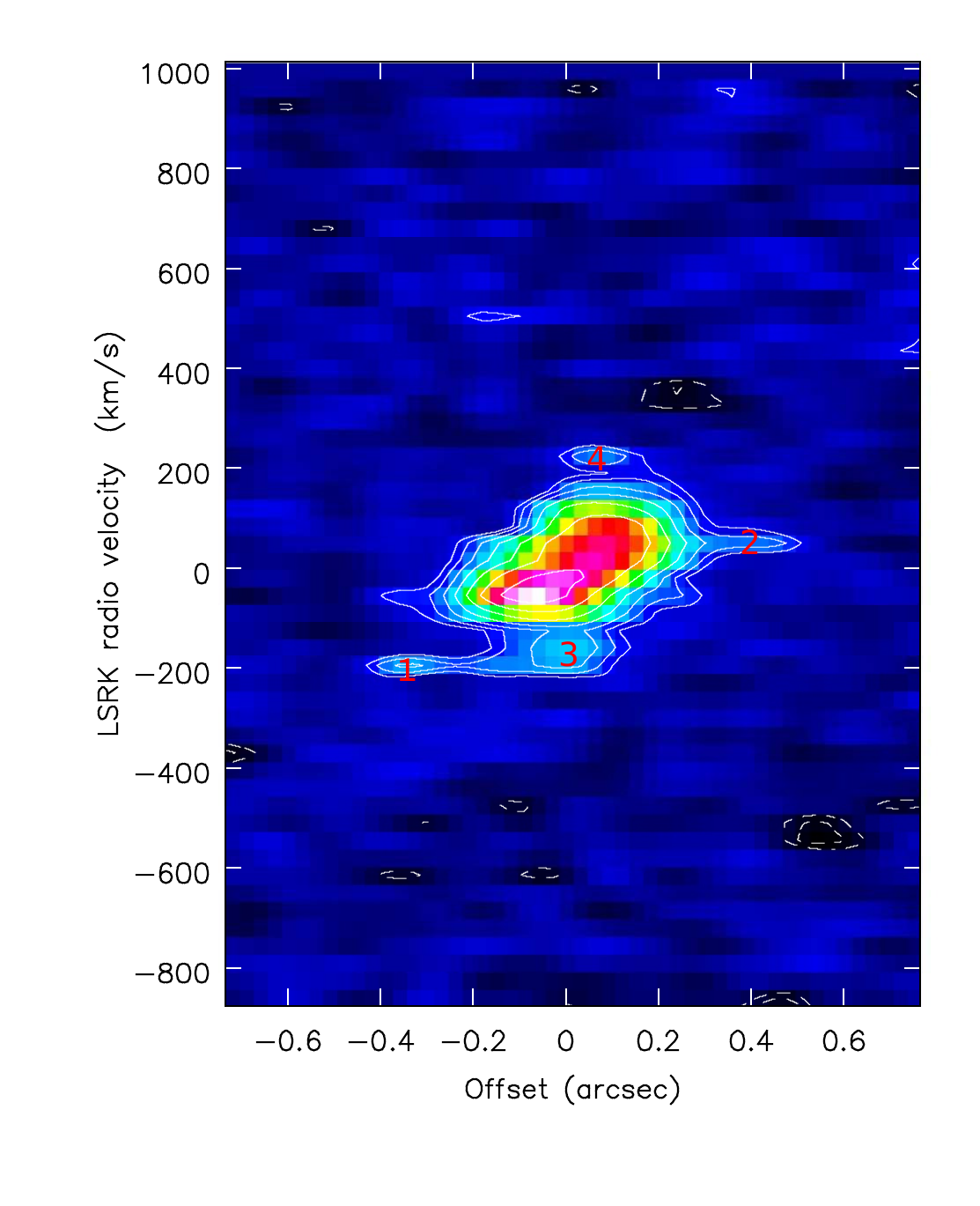}{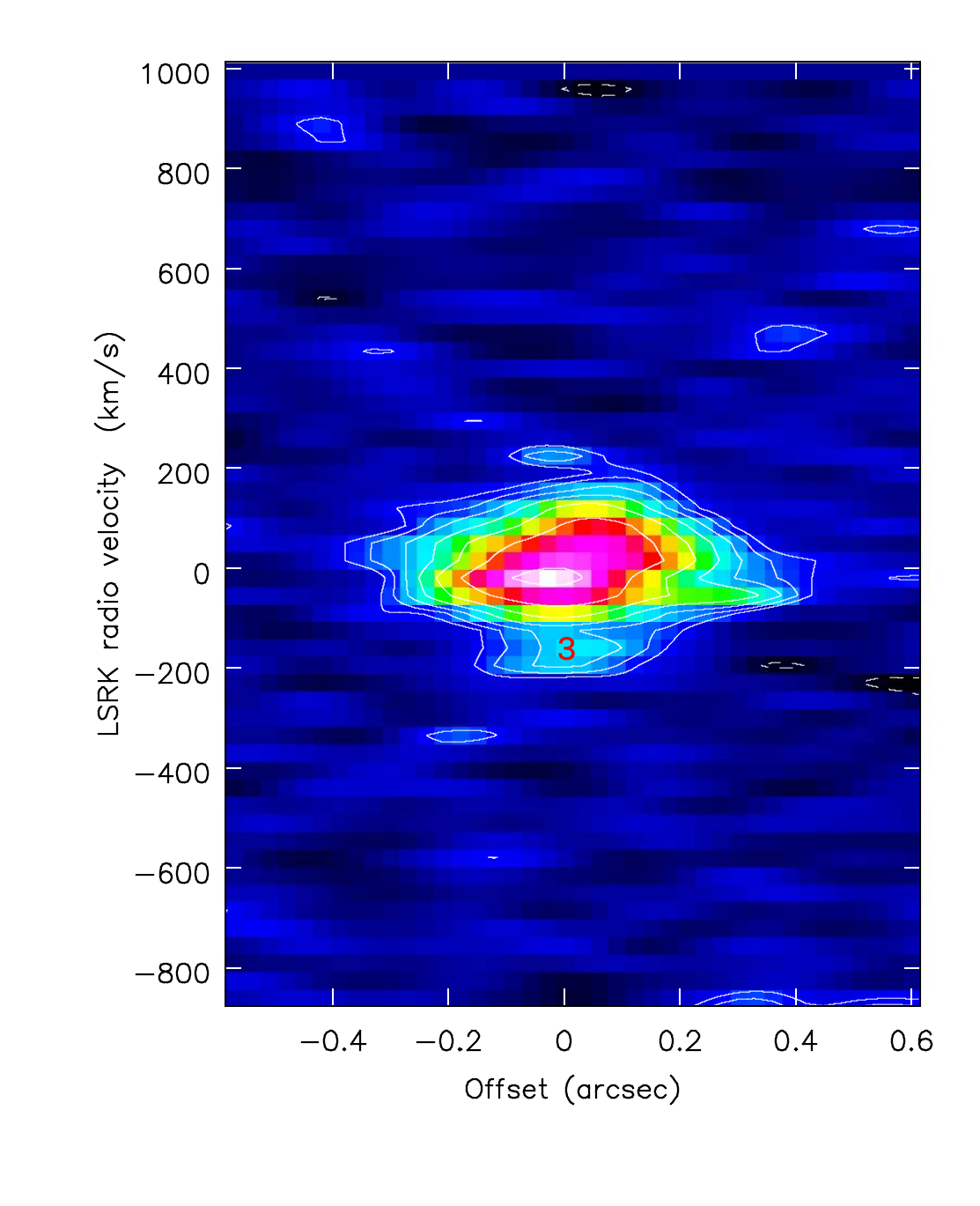}
\caption{\small The upper panels show the
intensity-weighted velocity and velocity dispersion maps of SDSS J0129$-$0035 (color maps). 
The maps are derived with the channel images including pixels at $\rm \geq3.5\sigma$. 
The black contours are the [C II] line intensity map (same as Figure 1). The numbers 1 and 2 mark two substructures
on the velocity map (see details in Section 3 and Section 4.2).
The lower panels show the Position-Velocity (PV) diagrams along the direction of the two
substructures (PA=127$^{\circ}$, left) and the perpendicular direction (PA=37$^{\circ}$, right).
The contour levels are [-2, 2, 2$\sqrt{2}$, 4, 4$\sqrt{2}$, 8, 8$\sqrt{2}$,
16]$\rm \times0.2\,mJy\,beam^{-1}$. The zero velocity corresponds to the [C II] redshift of z=5.7787 \citep{wang13}.
The central position for the PV diagrams is RA $\rm 01^{h}29^{m}58.513^{s}$ DEC $\rm -00^{d}35'39.82''$, and the
slit width is 0.09$''$.
There are also components in the PV-diagrams that show velocities of $\rm \pm200\,km\,s^{-1}$ within the 
central $\rm <0.2''$ region (see details in Section 3 and Section 4.2). We mark them as No. 3 and 4 in the PV diagrams. We also mark their position in the velocity map on the top-left panel. 
The velocity map and PV-diagrams suggest rotating motion of [C II]-emitting gas. 
The substructures (No. 1-4) may suggest additional clumps with complex kinematic properties.
}
\end{figure}

\begin{figure}
\includegraphics[height=2.5in]{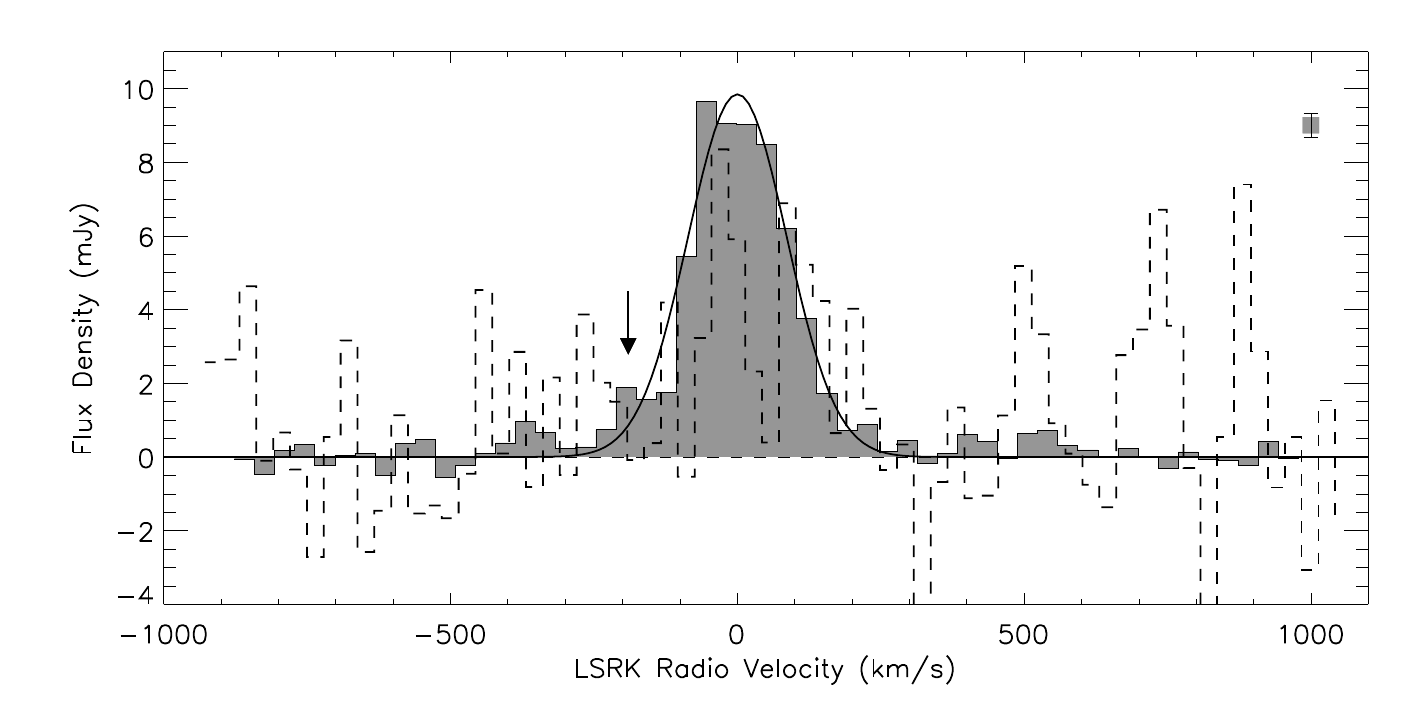}
\caption{The gray histogram represents the [C II] line spectrum of SDSS J0129-0035 integrated
over the line-emitting region within the 2$\sigma$ contour in the upper left panel of Figure 1.
The error bar on the top right denotes the typical 1$\sigma$ rms uncertainty per channel. The solid line 
shows a Gaussian line profile fitted to the [C II] line spectrum. The dashed line is the 
spectrum of the CO (6-5) line for this object scaled by a factor of 4 (Wang et al. 2011). 
The zero velocity corresponds to the [C II] redshift of z=5.7787 \citep{wang13}.
The arrow marks a tentative excess on the blue wing of the [C II] line spectrum compared to 
the single Gaussian line profile (see discussion in Section 3). }
\end{figure}

\begin{figure}
\includegraphics[height=2.8in]{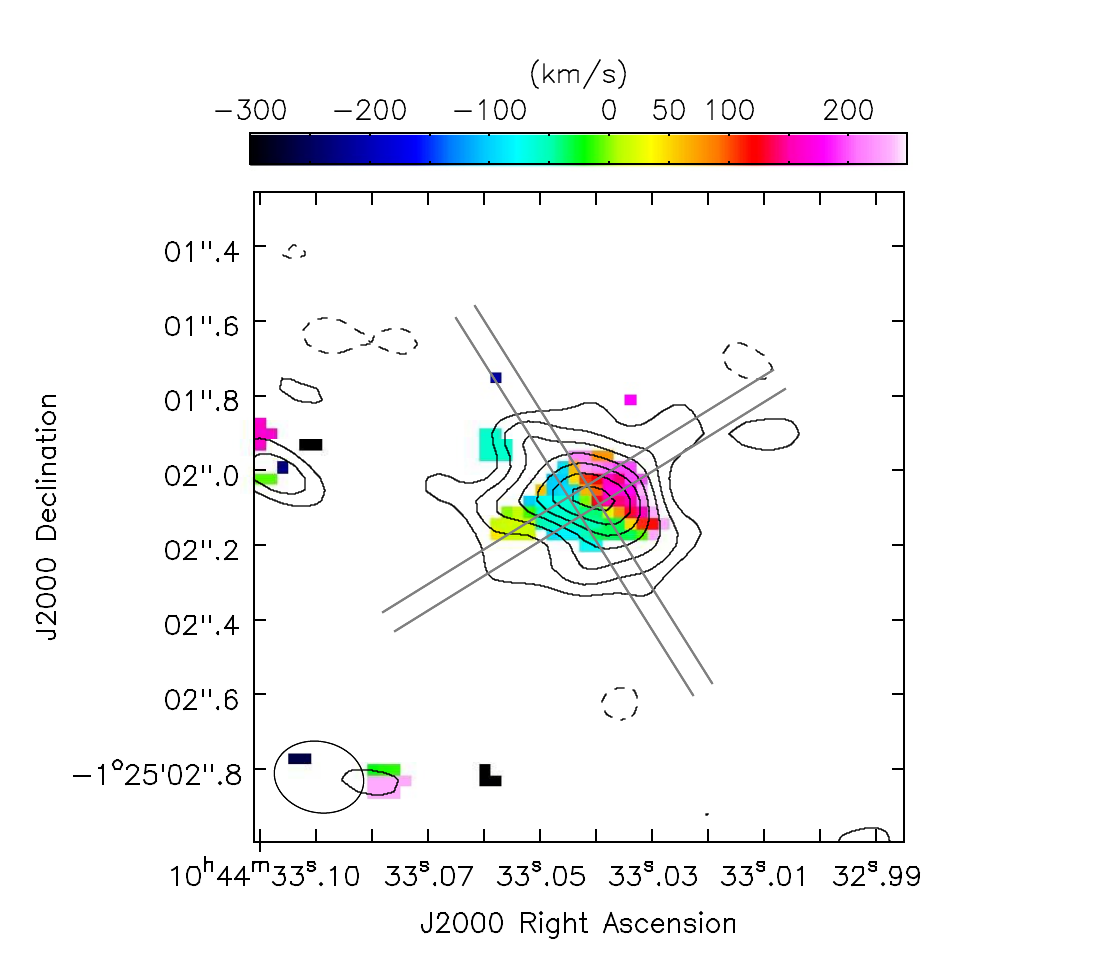}
\vskip -2.8in
\hspace*{3.0in}
\includegraphics[height=2.8in]{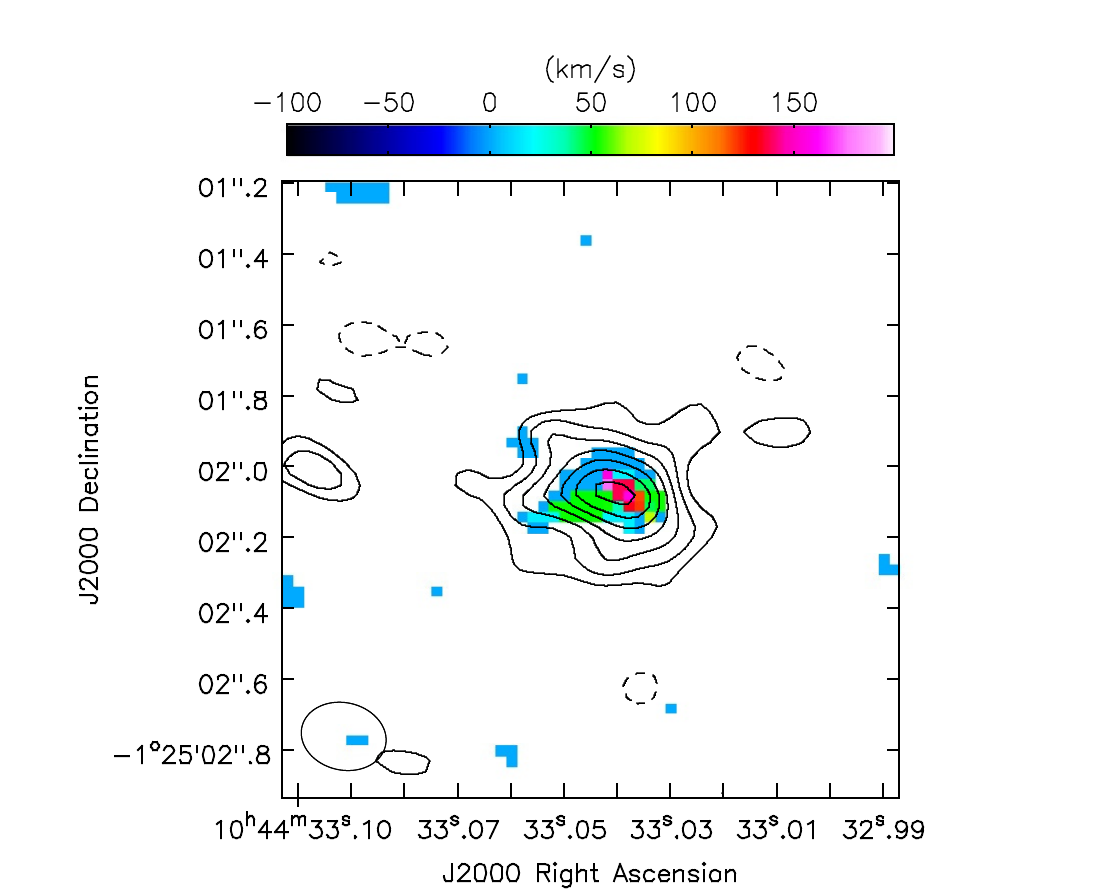}\\
\plottwo{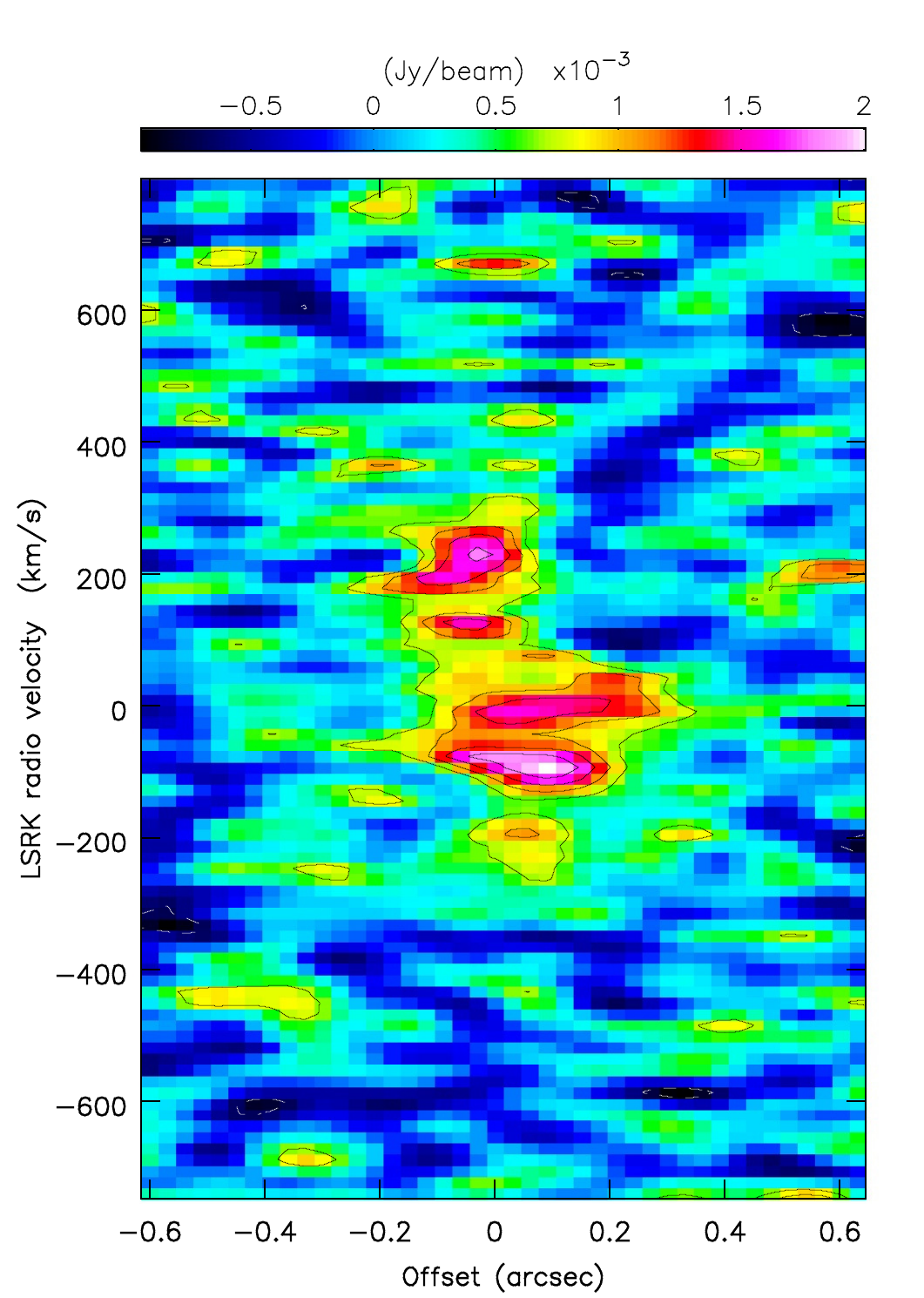}{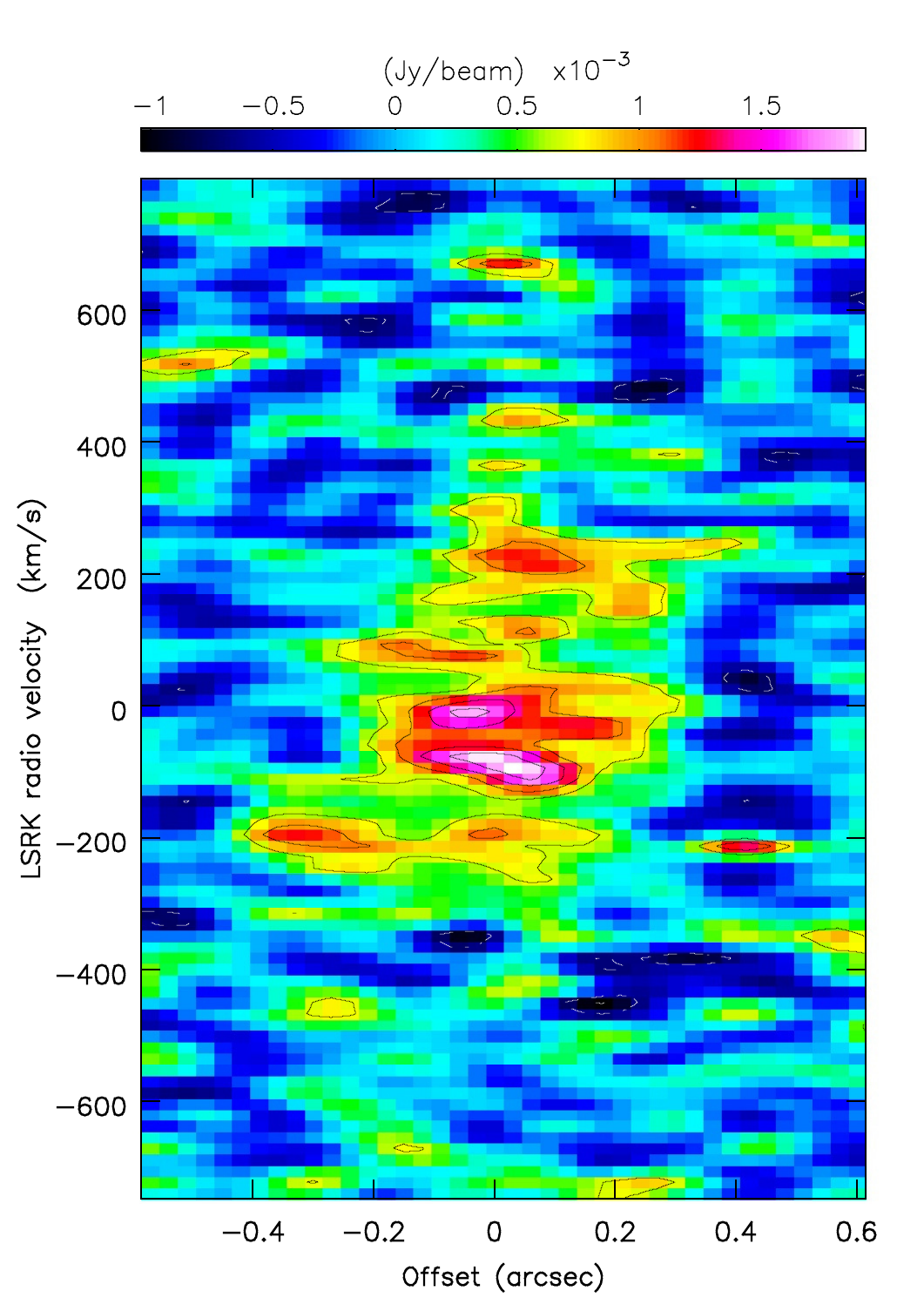}
\caption{The upper panels show the intensity-weighted velocity (left) and velocity dispersion (right) maps
of the [C II] line emission of SDSS J1044$-$0125, together with the line intensity map shown as contours (see Figure 4).
The velocity and velocity dispersion maps are calculated using only the $\rm >4\sigma$ pixels of the channel image.
The lower panels show the Position-Velocity (PV) diagrams along the directions of
PA=122$^{\circ}$ (left) and PA=32$^{\circ}$ (right, marked as gray lines on the velocity map on the upper left panel).
The contour levels are [-2, 2, 3, 4, 5]$\rm \times0.34\,mJy\,beam^{-1}$. The zero velocity corresponds 
to the [C II] redshift of z=5.7847 \citep{wang13}. The central position for the PV diagrams is RA $\rm 10^{h}44^{m}33.04^{s}$, DEC $\rm -01^{d}25'02.08''$, and the
slit width is 0.09$''$.}
\end{figure}

\begin{figure}
\includegraphics[height=1.9in]{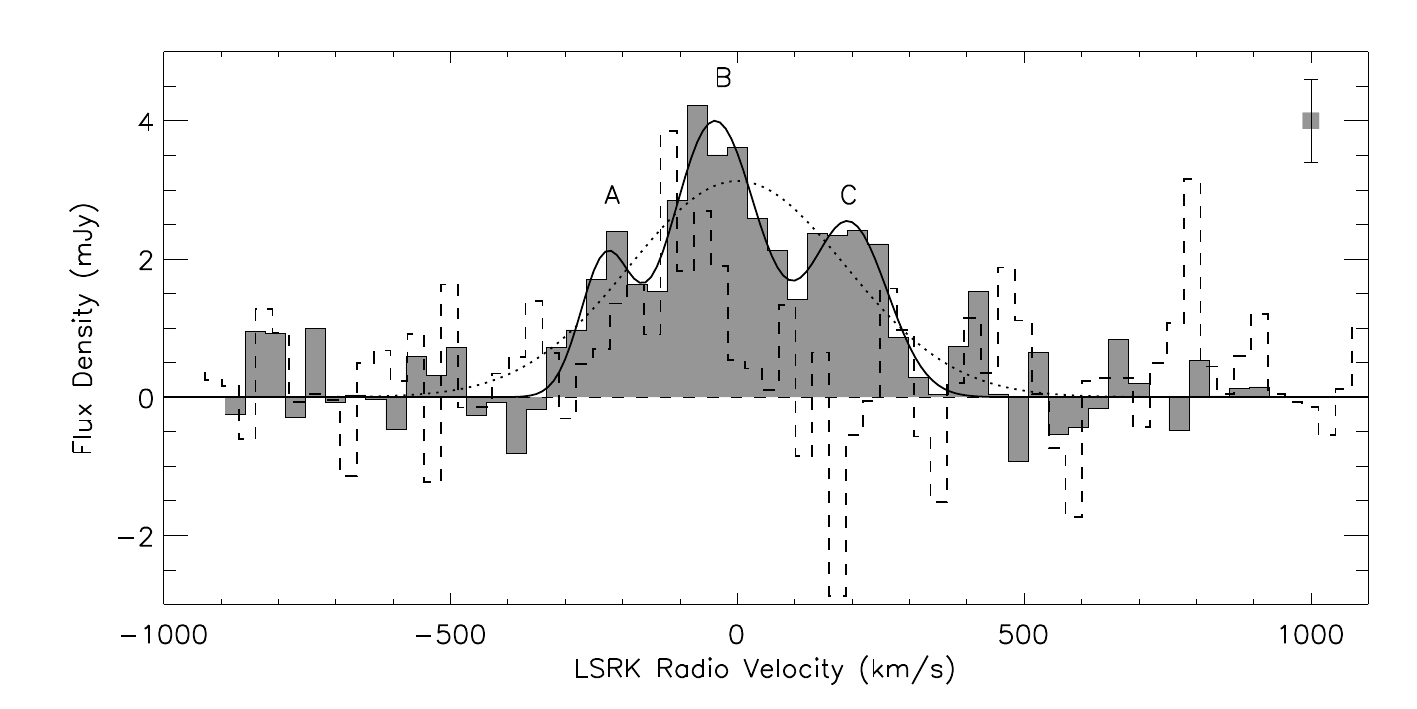}
\caption{[CII] line spectrum (gray histogram)
integrated over the line emitting region of SDSS J1044$-$0125 (i.e., within the 2$\sigma$ contour of 
the lower left panel of Figure 1). The error bar on top right represents the 1$\sigma$ rms for each channel. 
The dashed line is the spectrum of the CO (6-5) line
emission (Wang et al. 2011), scaled by a factor of 2. The zero velocity corresponds to the [C II]
redshift of 5.7847 \citep{wang13}. The dotted line show a single-Gaussian fitting to the [C II] line. The [C II] spectrum 
shows multiple peaks which cannot be well-fitted with a single-Gaussian 
component. Thus we also perform the fitting with three Gaussian components marked as A, B, and C in 
the spectrum and the solid line denotes the total emission of the three components (see Table 2).  
}
\end{figure}

\begin{figure}
\includegraphics[height=2.2in]{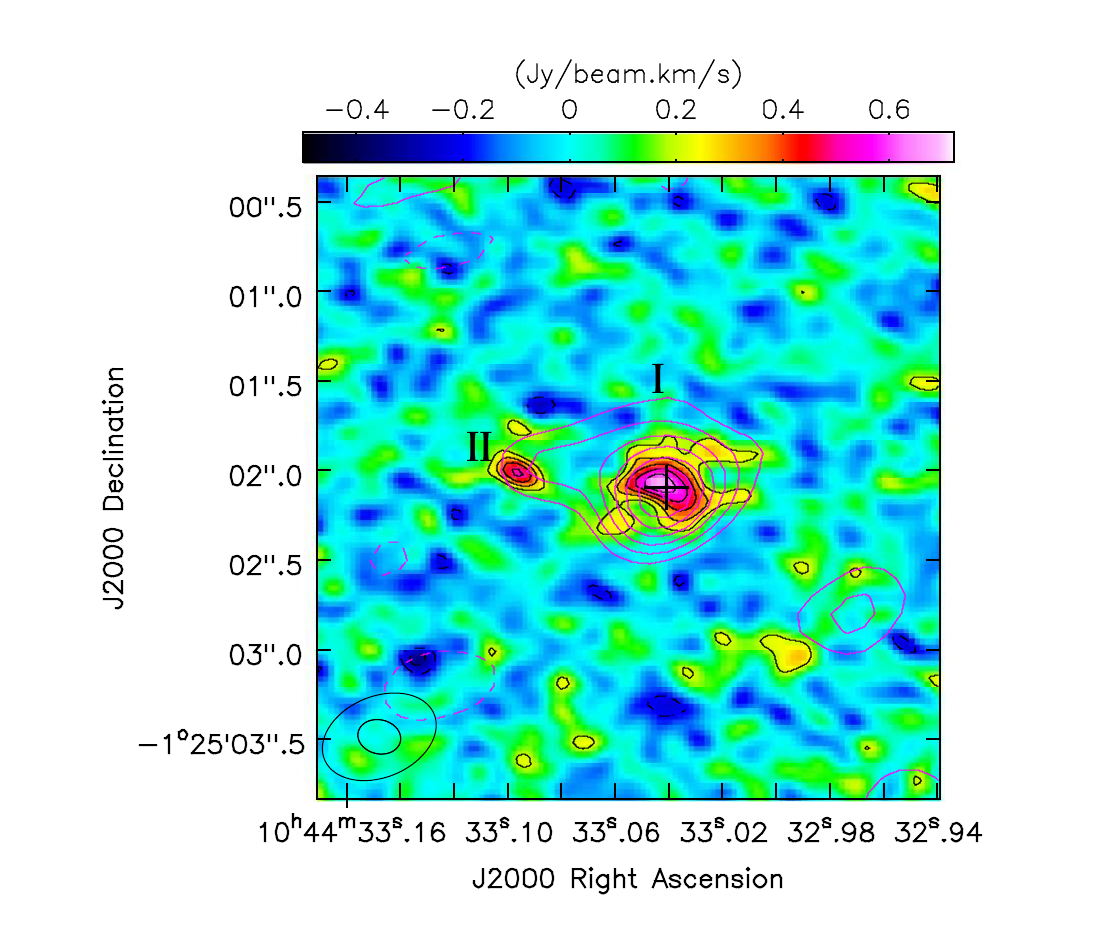}
\vskip -2.0in
\hspace*{2.2in}
\includegraphics[height=2.2in]{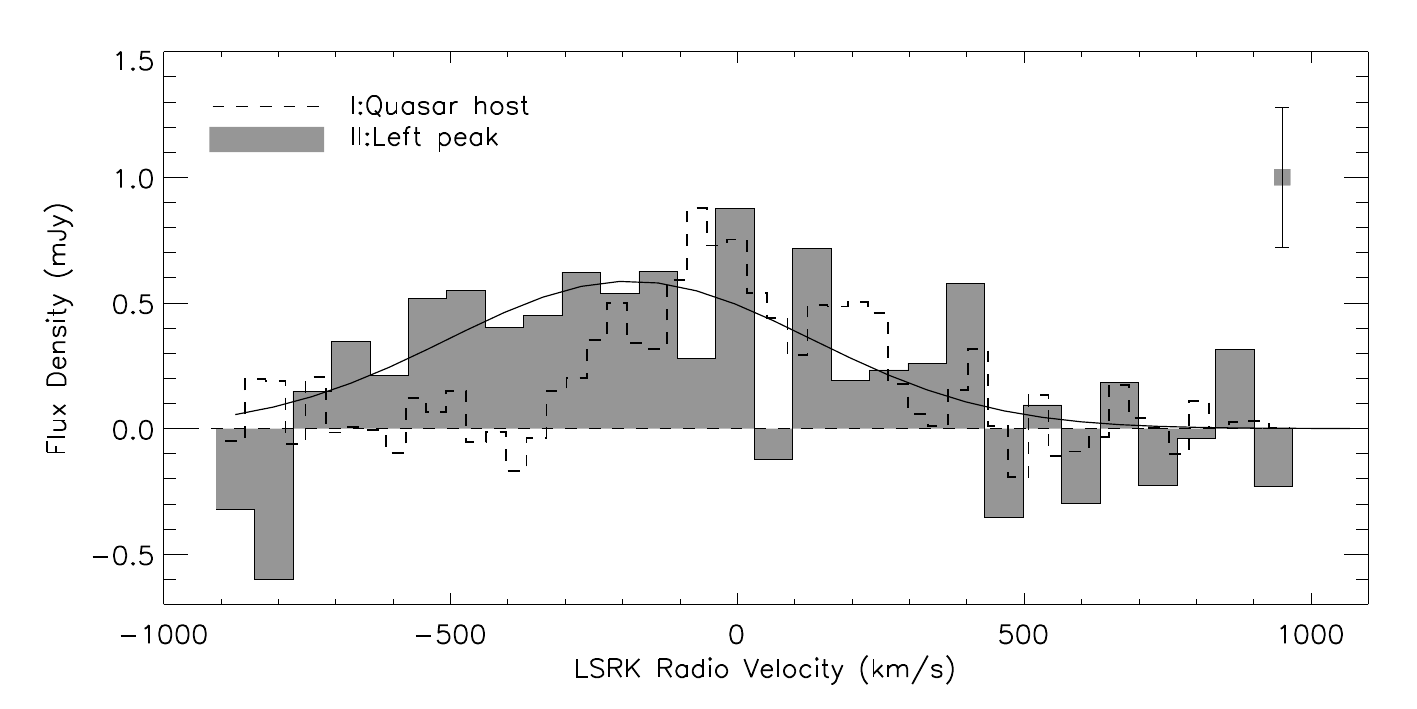}
\caption{The left panel is a [C II] intensity map  
integrated over the velocity range of $\rm -740\,km\,s^{-1}$ to $\rm 400\,km\,s^{-1}$, showing the 
component to the east of the quasar SDSS J1044-0125. We marks the [C II] emission from the quasar host 
galaxy as comonent 'I' and the east peak as component 'II'. The cross denotes the optical quasar position.
The black contours are [-3, -2, 2, 3, 4, 5, 6]$\rm \times0.1\,Jy\,beam^{-1}\,km\,s^{-1}$. 
The red contours are the low resolution ALMA Cycle 0 data of the [C II] velocity-integrated map published in Wang et al. (2013) 
with contour levels of [-2, 2, 3, 4, 5, 6, 7, 8]$\rm \times0.14\,Jy\,beam^{-1}\,km\,s^{-1}$.
The low resolution (FWHM beam size of $\rm 0''.66\times0''.45$) data
show some extension to this East peak. The solid line with gray fill in the 
right panel is the spectrum of the east peak, binned to a channel 
width of $\rm 67\,km\,s^{-1}$. The typical 1$\sigma$ rms noise of the spectrum is
shown in the top right, and the solid line is the Gaussian line profile fit to the data. 
The dashed line shows the [C II] spectrum of the quasar (the same as Figure 6), scaled to the same 
peak level for comparison. 
}
\end{figure}

\begin{figure}
\includegraphics[height=1.9in]{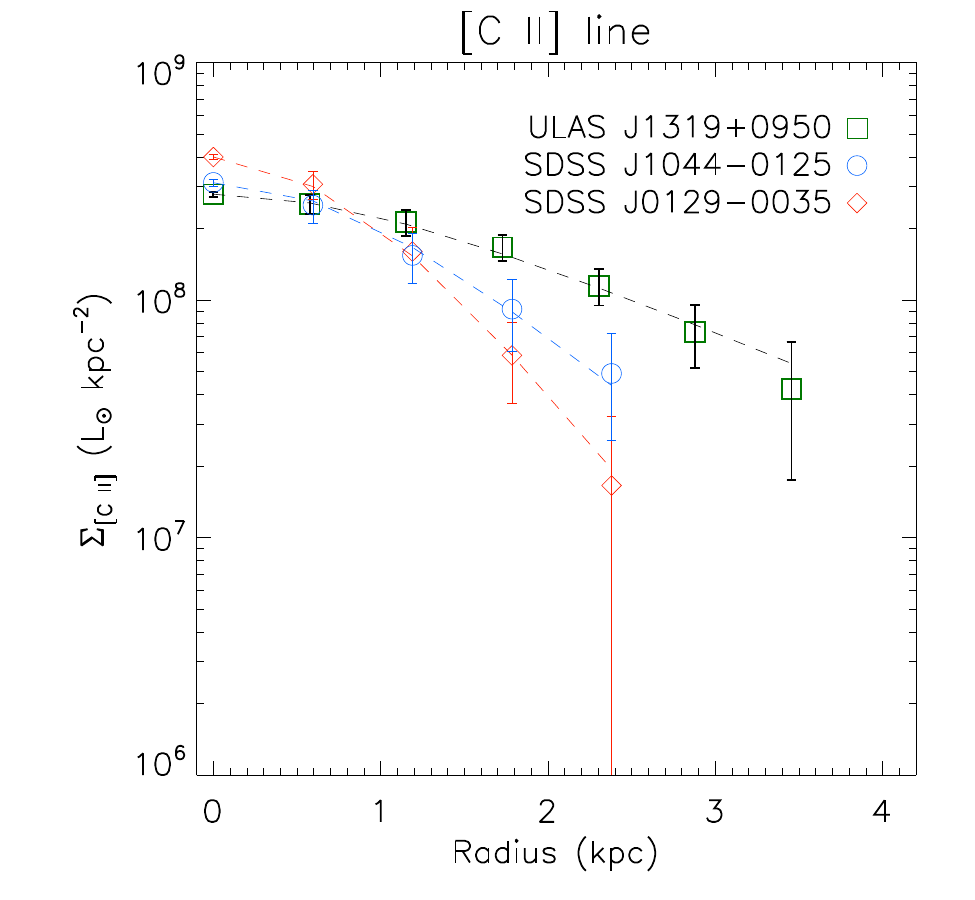}
\vskip -1.9in
\hspace*{2.0in}
\includegraphics[height=1.9in]{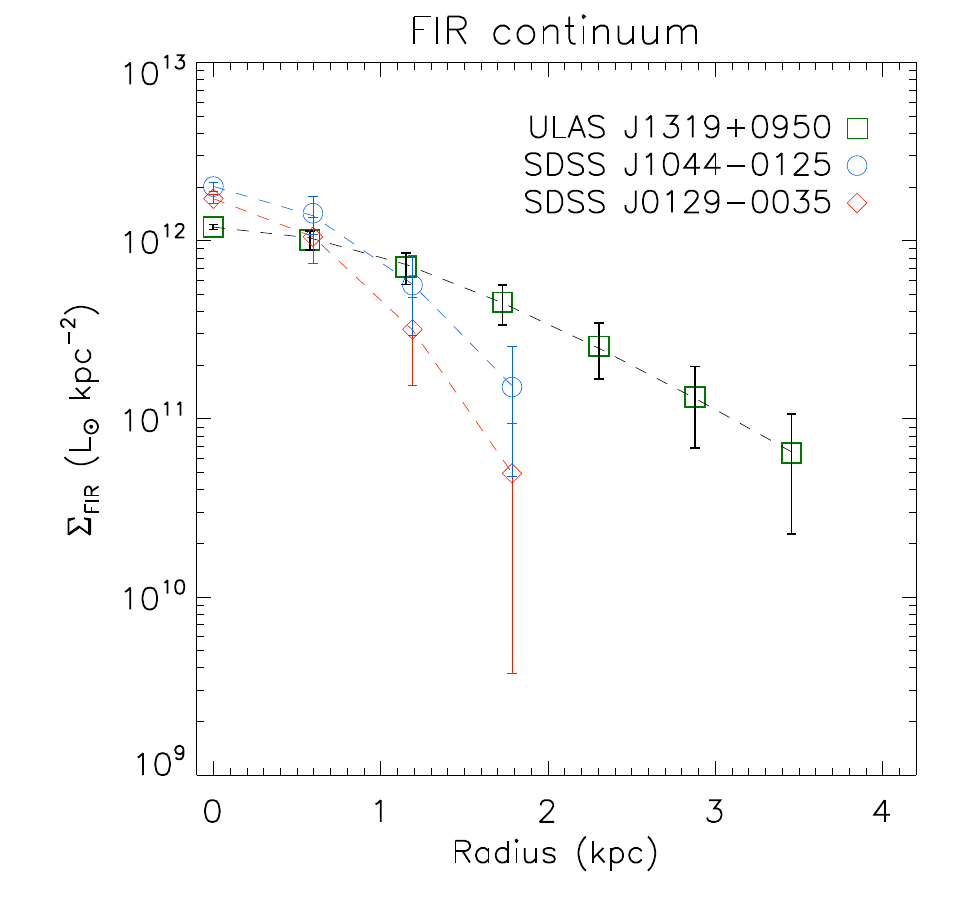}
\vskip -1.9in
\hspace*{4.0in}
\includegraphics[height=1.9in]{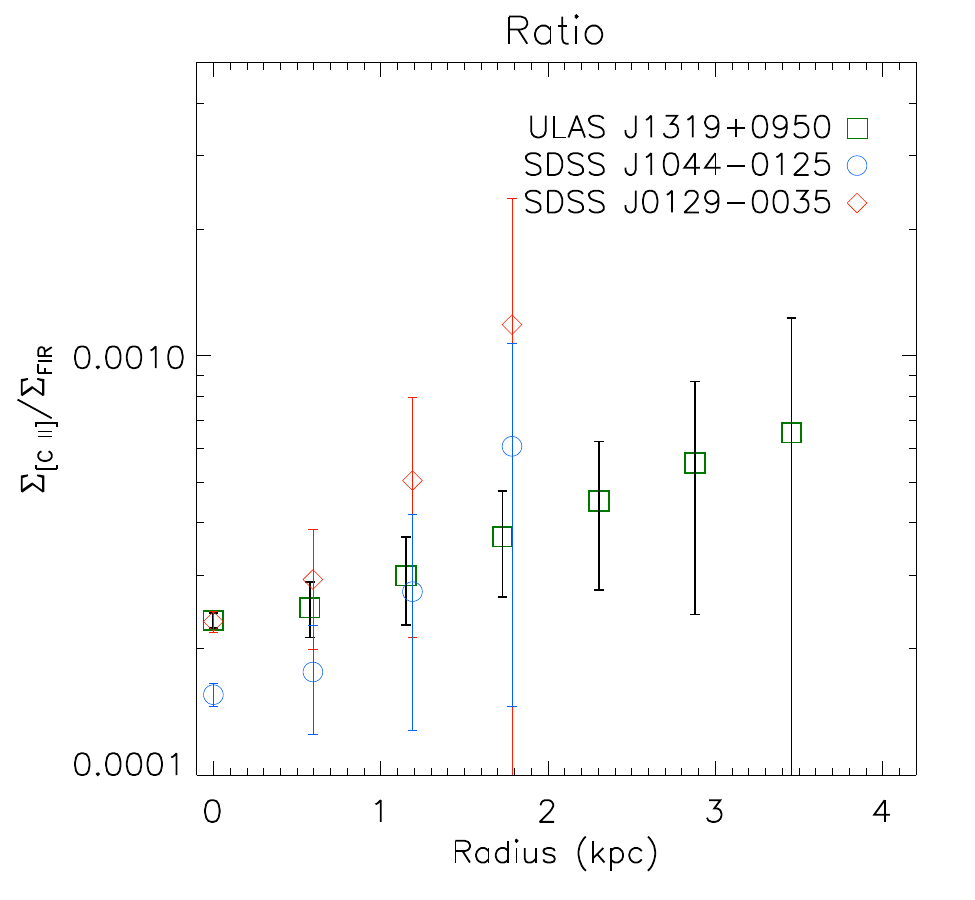}
\caption{Surface brightnesses of the [C II], dust continuum, and the [C II]-to-FIR surface
brightness ratios as a function of distance from the center of three high redshift quasars, ULAS J1319+0950 \citep{shao17}, SDSS J0129$-$0035, and SDSS J1044$-$0125 (this work) that have been observed
with ALMA at $\rm 0''.2\sim0''.3$ resolution. 
We calculate the mean surface brightnesses within the rings at central radii of 0$''$, 
0.1$''$, 0.2$''$, 0.3$''$, 0.4$''$, 0.5$''$, 0.6$''$ and a ring width of 0.1$''$. 
The scale of 0.1$''$ corresponds to a physical size of 0.58 kpc for ULAS J1319+0950, and 0.60 kpc for SDSS J0129$-$0035 and SDSS J1044$-$0125.
The error bars represent the standard deviation of the pixel values in each ring. The dashed lines 
represent an exponential disk model, convolved with the beam and 
fitted to the data. 
}
\end{figure}

\begin{figure}
\plotone{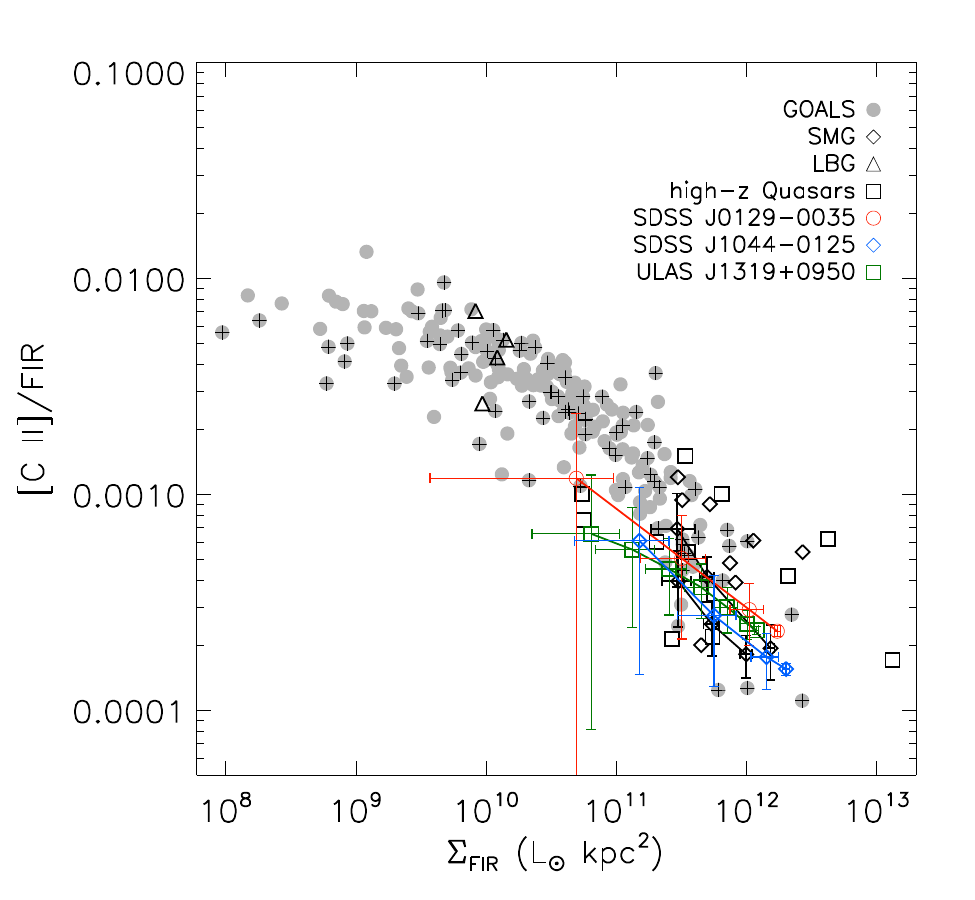}
\caption{\small [C II]-FIR emission ratio vs. FIR surface brightness of the two z=5.78 quasars 
resolved by ALMA in this work. We also include ULAS J1319+0950 from \citet{shao17} which 
was resolved with ALMA at a similar resolution. We compare these FIR luminous z$\sim$6 
quasar hosts to other systems, including the GOALS 
sample of local luminous infrared galaxies \citep{diazsantos13,diazsantos17,shangguan19}, the high-redshift 
submillimeter galaxies (SMGs, \citealp{riechers13,neri14,gullberg18,rybak19}), Lyman Break 
Galaxies (LBG, \citealp{capak15,jones17,hashimoto18}), and the z$\sim$6 to 7 
quasars \citep{wang13,venemans16,venemans17c,diazsantos16,decarli18}. 
For the two z=5.78 quasars in this paper, ULAS J1319+0950, and the two compact 
SMG in \citet{rybak19} that have resolved measurements, we plot measurements at 
different radii (see Figure 8 above and Figure 8 in \citealp{rybak19}), connected by a line.  
For the GOALS sample, we use the [C II]/FIR ratio extract from the $\rm 9.4''\times9.4''$ region of one PACS detector 
element where the [C II] and continuum emission peaks, see details in \citealp{diazsantos13,diazsantos17}. 
The plus sign indicates GOALS sources with possible AGN activity from emission line diagnostics, 
WISE colors, or SED decomposition \citep{shangguan19}. }
\end{figure}

\begin{figure}
\includegraphics[height=1.9in]{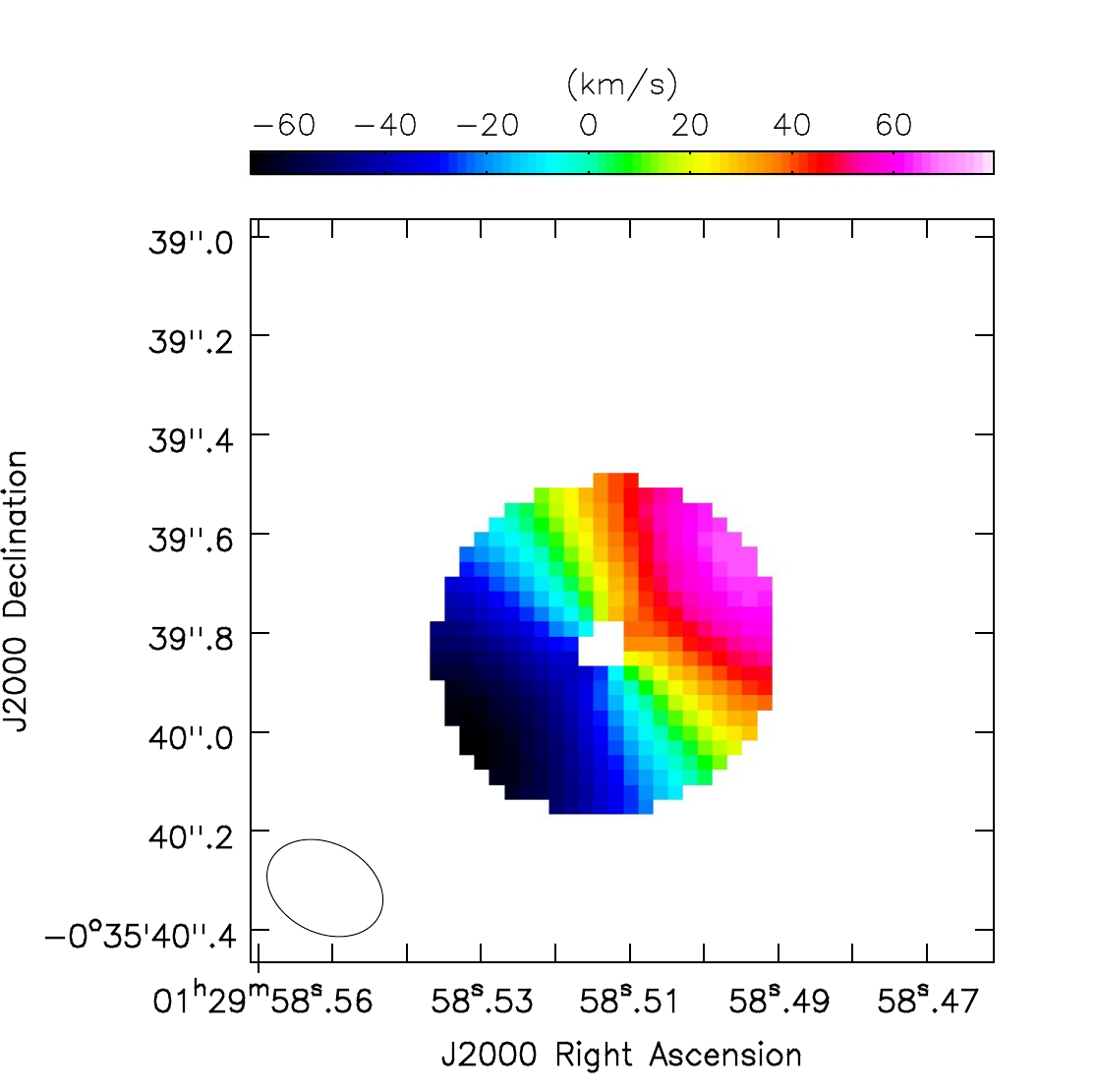}
\vskip -1.9in
\hspace*{2.0in}
\includegraphics[height=1.9in]{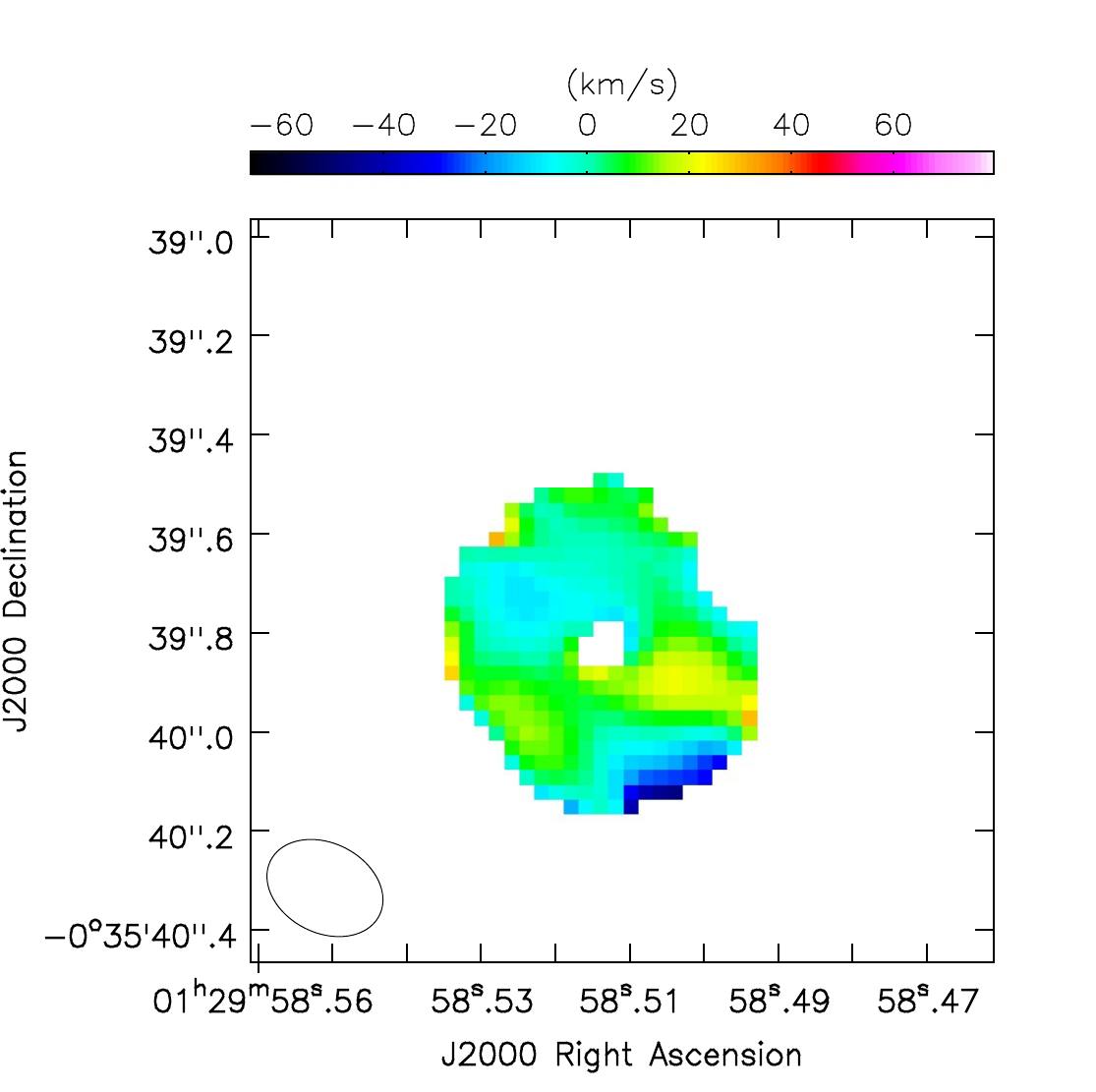}
\vskip -1.9in
\hspace*{4.0in}
\includegraphics[height=1.9in]{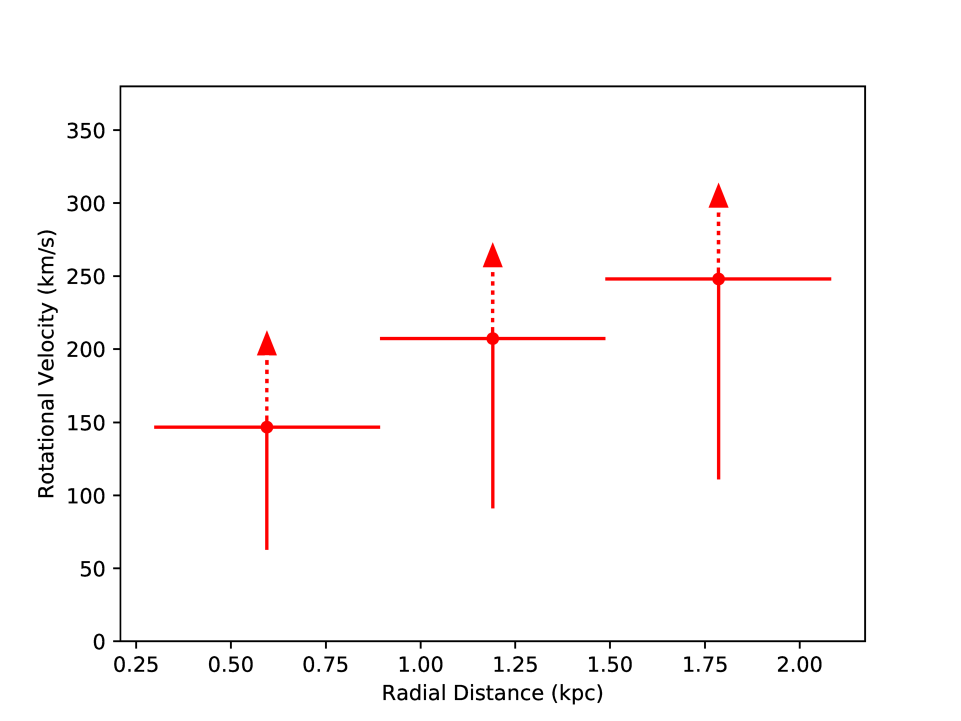}
\caption{Model of the velocity field of SDSS J0129$-$0035. The left panel shows the model of
velocity map with the fitting parameters described in Section 4.3. The middle panel shows the residual between the model
and the observed velocity field. The two outlying components marked as white stars in Figure 3 are not
included in the fit. The right panel shows the rotation velocity at radius
of 0$''$.1, 0$''$.2, and 0$''$.3 (i.e., 0.6 kpc, 1.2 kpc, and 1.8 kpc, at the quasar redshift). The inclination angle of the disk
is fitted to be $\rm incl=16^{\circ}$ with an uncertainty of 20$^{\circ}$. The lower limits of the rotation velocities correspond to
an inclination angle of $\rm 36^{\circ}$, while the upper limits of the velocities cannot be constrained
in the case of $\rm incl=0^{\circ}$.}
\end{figure}

\begin{figure}
\plotone{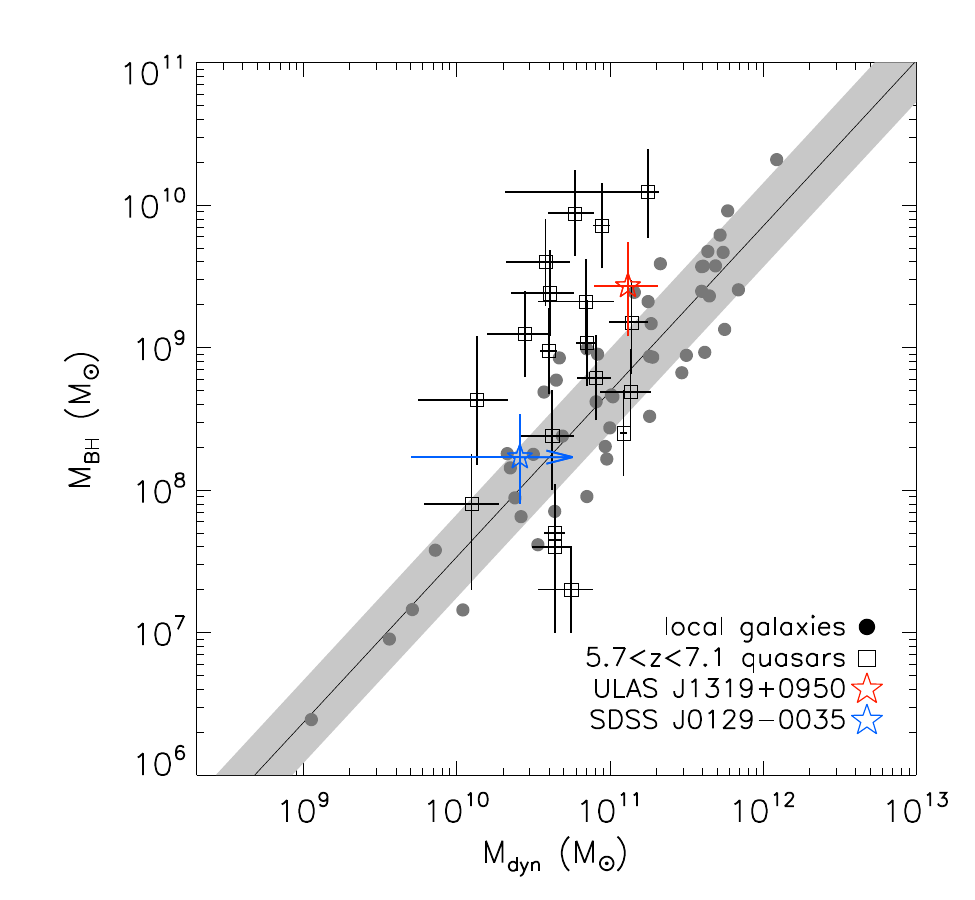}
\caption{\small SMBH mass vs. host galaxy dynamical mass of the z$\sim$6 to 7 quasars compared to the local galaxies. 
The black squares are the z$\sim$6 to 7 quasars from the literatures \citep{wang13,wang16,cicone15,willott15,willott17,venemans16,
feruglio18,izumi18,decarli18,wang19}. These objects all have source size measurements 
from [C II] or CO observations and most of then have inclination angles estimated from the major 
and minor axis ratios (see details in Section 4.3). Two of these z$\sim$6 quasars, SDSS J0129$-$0035 (this work) 
and ULAS J1319+0950 \citep{shao17} have [C II] emission resovled with ALMA at $\rm 0.2''\sim0.3''$, showing clear 
velocity gradients. Disk dynamical modeling and fitting of inclination angle are performed based on the 
velocity maps. Based on this we obtained a more reliable measurements of the host galaxy dynamical mass 
for ULAS J1319+0950 \citep{shao17} and a lower limit for SDSS J0129$-$0035. We plot ULAS J1319+0950 
and SDSS J0129$-$0035 as a red and blue stars, respectively.  
The gray dots denote the sample of local galaxies and the solid 
line and the gray area represent the relationship of the 
local galaxies with $\rm \pm0.3dex$ intrinsic scatter \citep{kormendy13}. }
\end{figure}

\begin{table}
\caption{Line and continuum parameters}
{\scriptsize
\begin{tabular}{lcc}
\hline \noalign{\smallskip}
\hline \noalign{\smallskip}
    & SDSS J0129$-$0035& SDSS J1044$-$0125\\
\hline \noalign{\smallskip}
[C II] line flux ($\rm Jy\,km\,s^{-1}$)  & 2.11$\pm$0.09 & 1.52$\pm$0.15\\
Peak ($\rm Jy\,beam^{-1}\,km\,s^{-1}$) & 0.80$\pm$0.03 & 0.64$\pm$0.08 \\
$\rm FWHM_{[CII]}$ ($\rm km\,s^{-1}$)  & 200$\pm$8 & 440$\pm$40\\
$[$C II$]$ central position (J2000)  & $\rm 01^h29^m58.513^s\,-00^{\circ}35'39.83''$ & $\rm 10^h44^m33.042^s\,-01^{\circ}25'02.08''$ \\
Deconvolved FWHM [C II] size ({\em imfit})  & $\rm (0''.32\pm0''.03)\times(0''.27\pm0''.03)$ & $\rm (0''.41\pm0''.07)\times(0''.30\pm0''.06)$\\
FWHM [C II] size ({\em uvmodelfit}) & $\rm (0''.32\pm0''.01)\times(0''.26\pm0''.01)$ & $\rm (0''.32\pm0''.02)\times(0''.26\pm0''.03)$ \\
$\rm L_{[CII]}(L_{\odot})$   & (1.96$\pm$0.08)$\times10^{9}$ & (1.41$\pm$0.14)$\times10^{9}$ \\
Continuum (mJy) & 2.82$\pm$0.04  & 3.02$\pm$0.11\\
Continuum central position (J2000)  & $\rm 01^h29^m58.513^s\,-00^{d}35'39.84''$& $\rm 10^h44^m33.040^s\,-01^{d}25'02.10''$ \\
Deconvolved FWHM continuum size ({\em imfit})  & $\rm (0''.18\pm0''.01)\times(0''.16\pm0''.01)$ & $\rm (0''.16\pm0''.02)\times(0''.15\pm0''.02)$ \\
FWHM continuum size ({\em uvmodelfit}) & $\rm (0''.172\pm0''.003)\times(0''.158\pm0''.005)$ & $\rm (0''.159\pm0''.005)\times(0''.154\pm0''.008)$ \\
Optical quasar position (J2000) & $\rm 01^h29^m58.524^s\,-00^{d}35'39.77''$& $\rm 10^h44^m33.041^s\,-01^{d}25'02.09''$ \\
Cycle 0 [C II] line flux$^{a}$ ($\rm Jy\,km\,s^{-1}$) & 1.99$\pm$0.12  &  1.70$\pm$0.30   \\
Cycle 0 Continuum (mJy) &  2.57$\pm$0.06  & 3.12$\pm$0.09 \\
\noalign{\smallskip} \hline
\end{tabular}\\
$^{a}$ Measurements of [C II] line flux and continuum flux density using the ALMA Cycle 0 data at 0.6$''$ 
resolution from Wang et al. (2013). The calibration uncertaintes are not included here, which is better 
than 15-20\% for the Cycle 0 data and better than 10\% for the new data presented here.
}
\end{table}

\begin{table}
\caption{Spectral fitting parameters for SDSS J1044-0125}
{\scriptsize
\begin{tabular}{lccc}
\hline \noalign{\smallskip}
\hline \noalign{\smallskip}
Component &  Center & FWHM & Flux \\
          &  $\rm km\,s^{-1}$ & $\rm km\,s^{-1}$ & $\rm Jy\,km\,s^{-1}$ \\
\hline \noalign{\smallskip}
A         &  -230$\pm$18 & 103$\pm$40 & 0.21$\pm$0.09 \\
B         &  -39$\pm$12  & 180$\pm$43 & 0.76$\pm$0.15 \\
C         &  194$\pm$19  & 160$\pm$44 & 0.42$\pm$0.11 \\
\noalign{\smallskip} \hline
\end{tabular}\\
}
\end{table}

\end{document}